\documentclass[12pt, draftclsnofoot, onecolumn]{IEEEtran}

\usepackage{float}
\usepackage{stackengine}
\usepackage{setspace}
\usepackage{amssymb,amsfonts,latexsym}
\usepackage{graphicx}
\usepackage{subfigure}
\usepackage{psfrag}
\usepackage{algorithmicx}
\usepackage{blindtext}
\usepackage{comment}
\usepackage{cite}
\usepackage{url}
\usepackage{times}
\usepackage{bbm}
\usepackage{epsfig}
\usepackage{multirow}
\usepackage{longtable}
\usepackage{amsfonts}
\usepackage{amssymb}%
\usepackage{color}
\usepackage{mathrsfs}
\usepackage{bm}
\usepackage{booktabs}
\usepackage{threeparttable}
\usepackage[cmex10]{amsmath}
\usepackage{amsmath}
\usepackage[english]{babel}
\usepackage{mathbbold}
\usepackage{mathtools, nccmath}
\usepackage[linesnumbered,ruled]{algorithm2e}
\makeatother
\usepackage[table]{xcolor}
\usepackage{makecell}
\usepackage{stackengine}

\usepackage{stackengine}

\usepackage{stfloats}

\newtheorem{remark}{Remark}

\ifCLASSINFOpdf
\else
\fi

\begin{document}
			\title{Split Learning over Wireless Networks: Parallel Design and Resource Management}
\author{\small Wen Wu,~\IEEEmembership{\small Senior~Member,~IEEE,}
	Mushu Li,~\IEEEmembership{\small Member,~IEEE,}	  
	Kaige  Qu,~\IEEEmembership{\small Member,~IEEE,}\\
	Conghao Zhou,~\IEEEmembership{\small Student~Member,~IEEE,}
	Xuemin~(Sherman)~Shen,~\IEEEmembership{\small Fellow,~IEEE}
		Weihua~Zhuang,~\IEEEmembership{\small Fellow,~IEEE},\\
				Xu~Li, and Weisen~Shi
		\thanks{W. Wu is with the Frontier Research Center, Peng Cheng Laboratory, Shenzhen,  China, 518055 (email: wuw02@pcl.ac.cn);}
	\thanks{M. Li, K. Qu, C. Zhou,  X. Shen, and W. Zhuang are with the Department of Electrical and Computer Engineering, University of Waterloo, Waterloo, Ontario, N2L 3G1, Canada, email: \{m475li, k2qu, c89zhou, sshen, wzhuang\}@uwaterloo.ca; }
	\thanks{X. Li and W. Shi are with Huawei Technologies Canada Inc., Ottawa, Ontario, K2K 3J1, Canada, email: \{xu.lica, weisen.shi1\}@huawei.com.}
	}

%
\maketitle

\begin{abstract}

Split learning (SL)  is a collaborative learning framework, which can train an artificial intelligence (AI) model between a device and an edge server by splitting the AI model into a  device-side model and a server-side model at a cut layer. The existing SL approach conducts the  training process sequentially across devices, which incurs significant training latency especially when the number of devices is large. In this paper,  we design a novel SL scheme to reduce the training latency, named \underline{C}luster-based  \underline{P}arallel \underline{SL} (CPSL) which conducts model training in a ``first-parallel-then-sequential" manner. 
Specifically, the CPSL is to partition devices into several clusters, parallelly train device-side models in each cluster and aggregate them, and then sequentially train the whole AI model across clusters, thereby parallelizing the  training process and reducing training latency. Furthermore, we propose a resource management algorithm to minimize the training latency of CPSL considering device heterogeneity and network dynamics in wireless networks. This is achieved by stochastically optimizing the cut layer selection, real-time device clustering, and radio spectrum allocation. The proposed two-timescale algorithm can jointly make the cut layer selection decision in a large timescale and device clustering and radio spectrum allocation decisions in a small timescale. Extensive simulation results on non-independent and identically distributed data demonstrate that the proposed solutions can greatly reduce the training latency as compared with the existing SL benchmarks, while adapting to network dynamics.

\end{abstract}

\section{Introduction}\label{sec:introduction}

With the wide deployment of Internet of things (IoT) devices and advanced sensing technologies, mobile devices are generating an unprecedented amount of data every day. Leveraging such voluminous data, state-of-the-art artificial intelligence (AI) techniques, especially deep neural networks (DNNs), have facilitated tremendous progress across a wide range of mobile applications, such as audio recognition, image classification, and object detection~\cite{gunduz2019machine, chen2020seek}. However, collecting device data is difficult or sometimes impossible because privacy laws and regulations shelter device data~\cite{regulation2018general}. Distributed learning frameworks, e.g., federated learning (FL), train AI models without sharing device data, such that  data privacy can be preserved~\cite{bonawitz2019towards}. In FL, devices parallelly train a shared AI model on their respective local dataset and  upload only the shared model parameters to the edge server. However, FL suffers from significant communication overhead since large-size AI models are uploaded and from prohibitive device computation workload since the computation-intensive  training process is only conducted at devices.

\begin{figure}[t]
	\vspace{-0.3cm}
	\centering
	\renewcommand{\figurename}{Fig.}	
	\begin{subfigure}[]{
			\label{fig:split_learning}
			\includegraphics[width=0.4\textwidth]{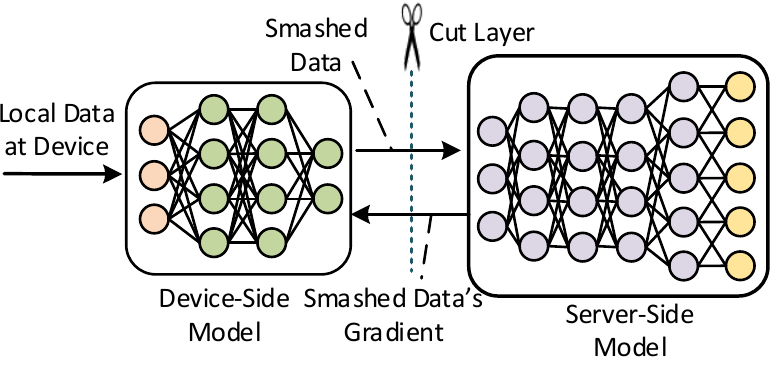}}
	\end{subfigure}
	~
	\begin{subfigure}[]{
			\label{fig:workload_communicaiton_vs}
			\includegraphics[width=0.3\textwidth]{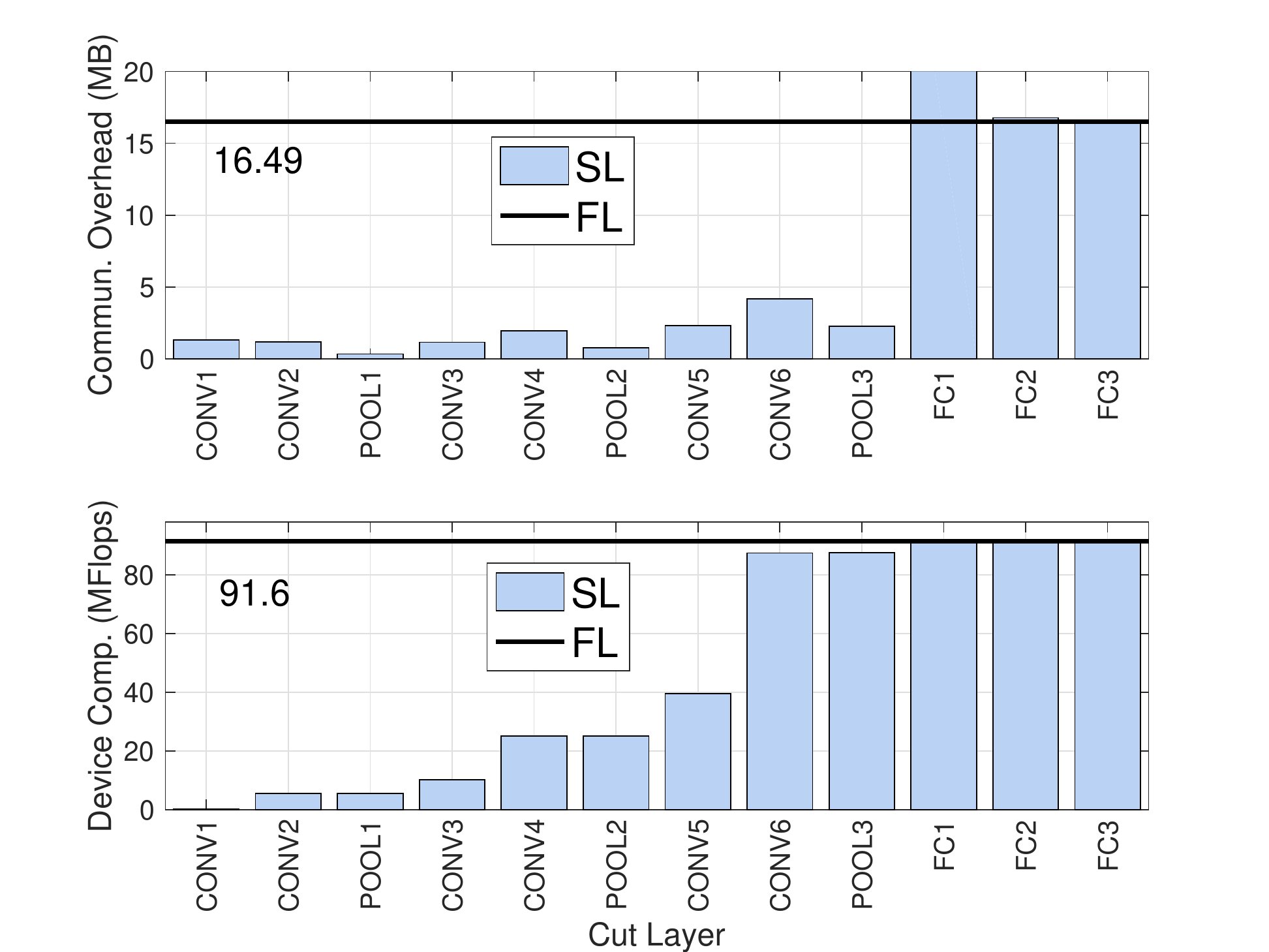}}
	\end{subfigure}
	\caption{(a) SL splits the whole AI model into a device-side model (the first four layers) and a server-side model (the last six layers) at a cut layer; and (b) the communication overhead and device computation workload of SL with different cut layers are presented in a LeNet example. }
	\vspace{-1cm}
\end{figure}

Split learning (SL), as an emerging collaborative learning framework, can effectively address the above issues. As shown in Fig.~\ref{fig:split_learning},  the basic idea of SL is to split an AI model at a \emph{cut layer} into a \emph{device-side model} running on the device and a \emph{server-side model} running on the edge server. The procedure of SL is as follows. {First}, the device executes the device-side model with local data and sends intermediate output associated with the cut layer, i.e., \emph{smashed data}, to the edge server, and then the edge server executes the server-side model, which completes the forward propagation (FP) process. {Second}, the edge server updates the server-side model and sends \emph{smashed data’s gradient} associated with the cut layer to the device, and then the device updates the device-side model, which completes the backward propagation (BP) process. In this way, SL for one device is completed. Next, the updated device-side model is transferred to the next device to repeat the above process until all the devices are trained. In SL, communication overhead is reduced since only small-size device-side models, smashed data, and smashed data's gradients are transferred. In addition, device computation workload is reduced since devices only train a part of the AI model. In the LeNet example shown in Fig.~\ref{fig:workload_communicaiton_vs}, compared with FL, SL with cut layer POOL1 reduces communication overhead by 97.8\% from 16.49\;MB to 0.35\;MB, and device computation workload by 93.9\% from 91.6\;MFlops to 5.6\;MFlops.  Due to its superior efficiency, SL is potentially suitable to resource-constrained IoT devices~\cite{gao2020end}. However, when multiple devices participate in SL, all the devices interact with the edge server in a \emph{sequential} manner, incurring significant training latency especially when the number of devices is large. To reduce the training latency, can we conduct the training process in a more efficient manner?






In this paper, we propose a novel low-latency SL scheme, named \underline{C}luster-based  \underline{P}arallel \underline{SL} (CPSL), which  parallelizes the device-side model training. At the beginning of the training process, all the devices are partitioned into several clusters, i.e., \emph{device clustering}.  The procedure of the CPSL operates in a ``first-parallel-then-sequential" manner, including: (1) \emph{intra-cluster learning} -  In each cluster, devices parallelly train respective device-side models based on local data, and the edge server trains the server-side model based on the concatenated smashed data from all the participating devices in the cluster. Then, the device-side models are uploaded to the edge server and aggregated into a new device-side model; and (2) \emph{inter-cluster learning} - The updated device-side model is transferred to the next cluster for intra-cluster learning. In this way, the AI model is trained in a sequential manner across clusters. In the CPSL, device-side models in each cluster are parallelly trained, which overcomes the sequential nature of SL and hence greatly reduces the training latency. We establish mathematical models to theoretically analyze the training latency of CPSL.

Furthermore, we propose a resource management algorithm to efficiently facilitate the CPSL  over wireless networks. Device heterogeneity and network dynamics lead to a significant straggler effect in CPSL, because the edge server requires the updates from all the participating devices in a cluster for server-side model training. To overcome this limitation, we investigate the resource management problem  in CPSL, which is formulated into a stochastic optimization problem to minimize the training latency by jointly optimizing cut layer selection, device clustering, and radio spectrum allocation. Due to the correlation among decision variables, network dynamics, and implicit objective function, the problem is difficult to solve. We  decompose the problem into two subproblems by exploiting the timescale separation of the decision variables, and then propose a two-timescale algorithm. Specifically, in the large timescale for the entire training process, a sample average approximation (SAA) algorithm is proposed to determine the optimal cut layer. In the small timescale for each training round, a joint device clustering and radio spectrum allocation algorithm is proposed based on the Gibbs sampling theory. Extensive simulation results on real-world non-independent and identically distributed (non-IID) data demonstrate that the newly proposed CPSL scheme with the corresponding resource management algorithm can greatly reduce training latency as compared with state-of-the-art SL benchmarks, while adapting to network dynamics. The main contributions of this paper are summarized as follows:

\begin{itemize}
	\item  We propose a novel low-latency CPSL scheme by introducing parallel model training. We analyze the training latency of the CPSL;
	\item We formulate  resource management  as a stochastic optimization problem to minimize the training latency considering network dynamics and device heterogeneity;
	\item We propose a two-timescale resource management algorithm to jointly determine cut layer selection, device clustering, and radio spectrum allocation. 
	
	
\end{itemize}

The remainder of this paper is organized as follows. Related works and system model are presented in Sections~\ref{sec:preliminary} and~\ref{sec:system_model}, respectively. The CPSL scheme is proposed in Section~\ref{sec:Parallel split learning scheme}, along with training latency analysis in Section~\ref{sec:Training Delay Analysis and Problem Formulation}. We formulate the resource management problem in Section~\ref{sec: problem formulation}, and the corresponding algorithm is presented in Section~\ref{sec:resource Management Algorithm}.  Simulation results are provided in Section~\ref{sec:Simulation Results}. Finally, Section~\ref{sec:conclusions} concludes this research.

\section{Related Work}\label{sec:preliminary}

Federated learning is arguably the most popular distributed learning method in recent years, which has been widely investigated. Extensive works are devoted to optimizing FL performance from different research directions, such as multi-tier FL framework design to accommodate a large number of devices~\cite{liu2020client, zhang2021optimizing}, and model aggregation and compression techniques to reduce communication overhead~\cite{yang2020federated, han2020adaptive}. More importantly, to facilitate FL over dynamic wireless networks, several pioneering works develop tailored resource allocation algorithms for FL considering communication link unreliability~\cite{zhang2021federated, chen2020joint} and energy efficiency~\cite{yang2020energy, mo2021energy}. We refer interested readers to recent comprehensive surveys on FL~\cite{chen2021distributed, kang2020reliable, shen2021holistic}.

\begin{table}[t]
	\scriptsize
	\centering
		\vspace{-0.5cm}
	\caption{{Summary of notations}.}
	\label{Tab:Variables_and_notations}
	\vspace{-0.5cm}
	\begin{tabular}{l l l l l}
		\hline
		\hline
		\textbf{{Notation}} & \textbf{{Description}}&	\textbf{{Notation}} & \textbf{{Description}} \\
		\hline
		{$\mathbf{A}$} & {Device clustering decision }&
		{$B$} & {Mini-batch size }\\
		{${D}\left(\cdot \right)$} & {Training latency function}&
		{$f_s$} & {Edge server computing capability}\\
		{$\mathbf{f}$} & {Device computing capabilities}&
		{$\mathbf{h}$} & {Device channel conditions}\\
		{$\mathcal{L}$} & {Set of local epochs}&
		{$L(\mathbf{w})$} & {Average loss function with model parameter $\mathbf{w}$}\\
		{$l\left(\mathbf{z}, {y}\right)$ } & {Loss function for  data sample $(\mathbf{x}, {y})$}&
		
		{$\mathcal{M}$} & {Set of clusters}\\
		{$\mathcal{N}$} & {Set of devices}&
		{$\mathcal{K}_m$} & {Set of devices in cluster $m$}\\
		{$G$} & {Number of iterations}&
		{$\mathcal{S}_{m,k}$} & {Smashed data of device $k$ in cluster $m$}\\
		{$\mathcal{T}$} & {Set of training rounds}&
		{$v, \mathcal{V} $} & {Cut layer and the set of cut layers}\\
		{$W$} & {Subcarrier bandwidth}&		
		{$\mathbf{w}$} & {Model parameter}\\
		{$\mathbf{w}_{m,k}^d$} & {Device-side model of device $k$  in cluster $m$}&
		{$\mathbf{w}_m^e$} & {Server-side model  in cluster $m$}\\
		{$\mathbf{x}_{m}$} & {Radio spectrum allocation decision in cluster $m$ }&
		{$\bar{\mathbf{w}}_m^d$} & {Aggregated device-side model parameter in cluster $m$}\\
		{$\eta_d$, $\eta_e$} &{Learning rates of device/server-side models}  &
		{$\xi_s$, $\xi_g$} &{Data sizes of smashed data and smashed data's gradient}  \\ 
		{$\xi_d$} & {Data size of the device-side model}&	{$\gamma_s^B$, $\gamma_s^F$}& {Server-side model's BP and FP computation workloads }
		\\
		{$\gamma_d^B$, $\gamma_d^F$}& {Device-side model's BP and FP computation workloads} &{$\kappa$ }&{Computing intensity of processing units}
	\\
		{$\delta$ }& {Smooth factor }&
		{$\nabla l\left(\mathbf{w}\right)$ }& {Loss function's gradient}\\		
		\hline
		\hline
	\end{tabular}
	\vspace{-0.2cm}
\end{table}


Different from FL, the research on SL is still in its infancy. The basic idea of SL (or collaborative DNN training) is first introduced in~\cite{gupta2018distributed}. For basic knowledge on SL, one can refer to a tutorial paper and references therein~\cite{thapa2020advancements}. Under some assumptions, SL is functionally equivalent to centralized learning on the aggregated datasets~\cite{gupta2018distributed}. Recently, due to its superior efficiency and simplicity, SL is gaining substantial interest from industry and academia. In industry, an SL framework is implemented in some open-source applications~\cite{OpenMined, opensource}, and relevant services are developed by  start-ups~\cite{Acuratio}.  In academia, there are a growing body of research works investigating SL. A line of works conducts empirical studies in different scenarios. Koda \emph{et. al} apply SL to depth-image based millimeter-wave received power prediction, in which a significant communication latency reduction gain is achieved~\cite{koda2020communication}. A few works apply SL in medical fields, such as X-ray image classification~\cite{poirot2019split}. Another work investigates SL performance in IoT devices~\cite{gao2020end}. Pasquini \emph{et. al} study  security issues in the SL framework~\cite{pasquini2021unleashing}. An early empirical work compares the performance of SL with FL in terms of communication overhead~\cite{singh2019detailed}. The preceding works have attested SL performance gain in various settings. Another line of works focuses on designing and optimizing SL schemes. An SL variant with two cut layers is proposed, in which the first and the last layers are kept at devices, thereby avoiding sharing both data samples and their labels~\cite{vepakomma2018split}. A pioneering work proposes an online learning algorithm to determine the optimal cut layer to minimize the training latency~\cite{zhang2021learning}. An extended work in~\cite{wang2021hivemind} studies a more complicated SL scheme with multiple cut layers, using a low-complexity algorithm to select the optimal set of cut layers. As most of the existing studies do not incorporate network dynamics in the channel conditions as well as device computing capabilities, they may fail to identify the optimal cut layer in the long-term training process. Moreover, while the above works can enhance SL performance, they focus on SL for one device and do not exploit any parallelism among multiple devices, thereby suffering from long training latency when multiple devices are considered.


Recently,  a few early research works are proposed to reduce training latency~\cite{thapa2020splitfed, turina2020combining, jeon2020privacy}. More prominently, a pioneering work combines the ideas of SL and FL to parallelize the training process~\cite{thapa2020splitfed}. This work deploys multiple server-side models to parallelize the training process at the edge server, which speeds up SL at the cost of abundant storage and memory resources at the edge server, especially when the number of devices is large. 
Different from the existing works, we focus on a parallel SL solution with only one shared server-side model for supporting a large number of devices. Furthermore, taking network dynamics and device heterogeneity into account, we propose a resource management algorithm to optimize the performance of the proposed solution over wireless networks.

\begin{figure}[t]
				\vspace{-0.7cm}
	\renewcommand{\figurename}{Fig.}
	\centering
	\includegraphics[width=0.5\textwidth]{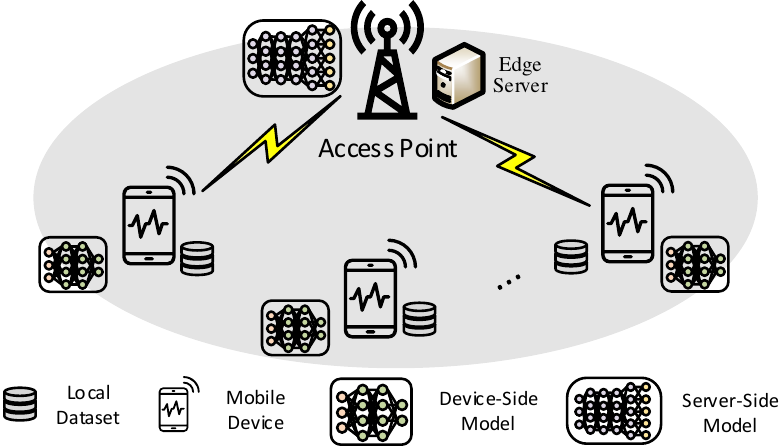}
	\caption{The SL framework in wireless networks. }
	\label{fig:system_model}
			\vspace{-0.7cm}
\end{figure}

\section{System Model}\label{sec:system_model}
As shown in Fig.~\ref{fig:system_model}, we consider a typical SL scenario over a wireless network comprising an access point (AP) and multiple devices. 
\begin{itemize}
	\item {AP}:  The AP is equipped with an edge server that can perform server-side model training. A server-side model, denoted by $\mathbf{w}^e$, is deployed at the AP. In addition, it is in charge of collecting network information, such as device computing capabilities and channel conditions, for making resource management decisions.
	\item {Device}: The set of devices is denoted by $\mathcal{N}=\{1,2,..., N\}$ where $N$ denotes the number of devices.  Each device is deployed with a device-side model, denoted by $\mathbf{w}^d$. The whole AI model is denoted by 
	\begin{equation}
		\mathbf{w}=\{\mathbf{w}^d; \mathbf{w}^e\}.
	\end{equation}
	The devices are endowed with computing capabilities, which can perform device-side model training. Each device possesses a local dataset, $\mathcal{D}_n= \{\mathbf{z}_i, y_i\}_{i=1}^{{D}_n}, \forall n \in \mathcal{N}$. Here, $\mathbf{z}_i \in \mathbb{R}^{Q\times 1}$ and $y_i\in \mathbb{R}^{1\times 1}$ represent an input data sample and its corresponding label, respectively, where $Q$ denotes the dimension of the input data sample. The aggregated dataset over all  devices is represented by $\mathcal{D}=\cup_{n =1}^N\mathcal{D}_n$. A summary of important notations in this paper is given in Table~\ref{Tab:Variables_and_notations}.
\end{itemize}




In  SL, the AP and devices collaboratively train the considered AI model without sharing the local data at devices. Let $l\left(\mathbf{z}_i, y_i; \mathbf{w}\right)$ represent the sample-wise loss function that quantifies the prediction error of   data sample $\mathbf{z}_i$ with regard to its label $y_i$ given model parameter  $\mathbf{w}$.\footnote{There are several types of loss functions in  model training, such as cross-entropy, mean squared error, and log likelihood~\cite{wang2019adaptive, mo2021energy}. In the simulation, the log likelihood loss function is adopted.}  The average loss function for device $n$ is given by $L_n(\mathbf{w})=\frac{1}{|\mathcal{D}_n|}\sum_{\{\mathbf{z}_i, y_i\} \in \mathcal{D}_n}l\left(\mathbf{z}_i, y_i; \mathbf{w}\right), \forall n\in \mathcal{N}.$ The global loss function, $L(\mathbf{w})$, is the average with weights proportional to the number of data samples in each dataset, given by
\begin{equation}
	L(\mathbf{w})=\frac{\sum_{n\in \mathcal{N}}|\mathcal{D}_n| L_n(\mathbf{w}) }{\sum_{n\in \mathcal{N}}|\mathcal{D}_n|}.
\end{equation}
The problem of SL boils down to identifying optimal model parameter $\mathbf{w}^\star$ with minimum global loss $\mathbf{w}^\star =\arg\min_{\mathbf{w}} L(\mathbf{w})$.

To minimize the global loss, the model parameter is  \emph{sequentially} trained across devices in the vanilla SL scheme, i.e., conducting model training with one device and then moving to another device, as shown in Fig.~\ref{fig:vanilla}. %
Sequentially training behaviour  may incur significant training latency since it is proportional to the number of devices, especially when the number of participating devices is large and device computing capabilities are limited. Such limitation motivates the following design of a parallel version of SL for training latency reduction.


%



\section{CPSL Scheme Design}\label{sec:Parallel split learning scheme}
\begin{figure}[t]
	\centering
	\renewcommand{\figurename}{Fig.}	
	\begin{subfigure}[Vanilla SL]{
			\label{fig:vanilla}
			\includegraphics[width=0.5\textwidth]{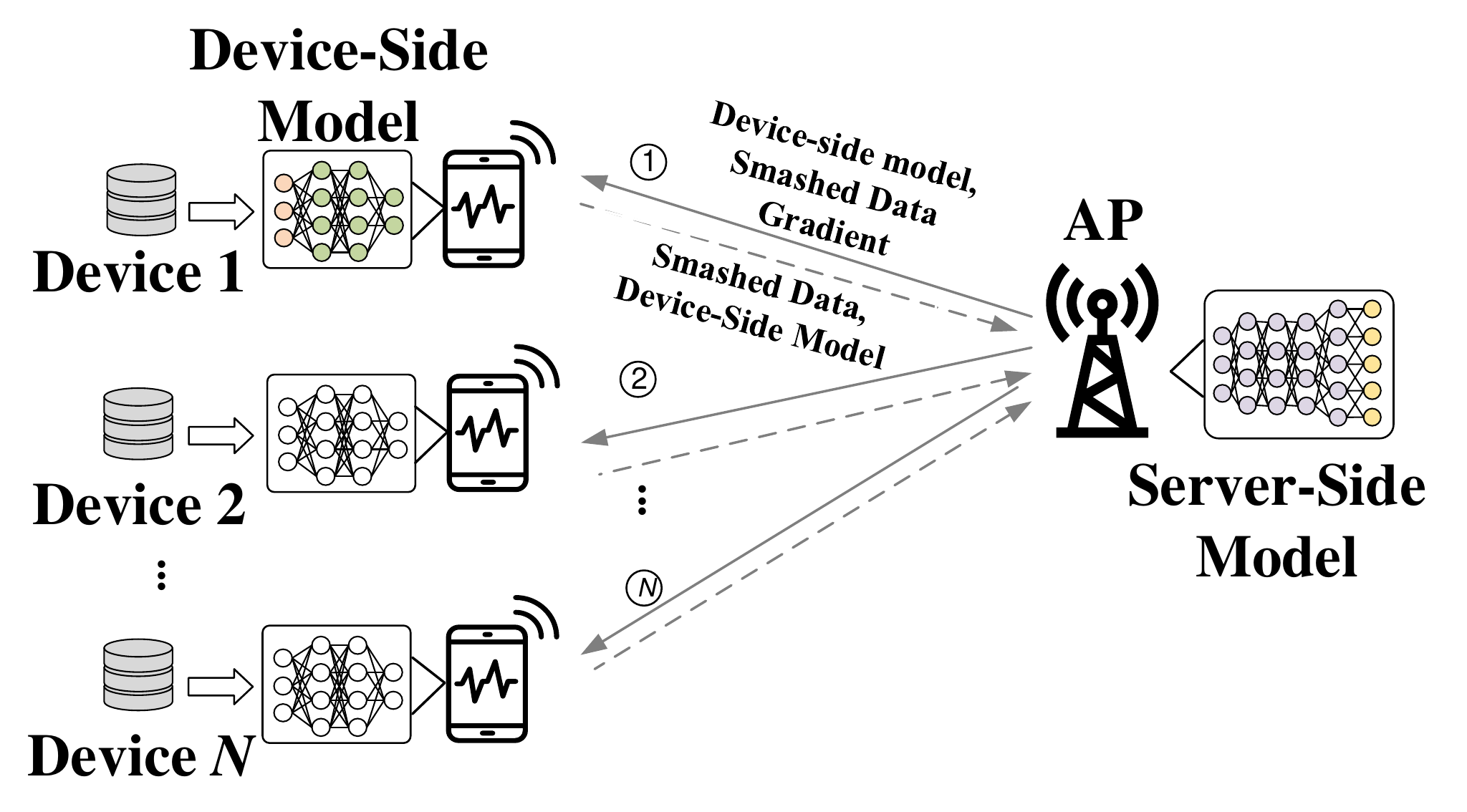}}
	\end{subfigure}
	~
	\begin{subfigure}[CPSL]{
			\label{fig:CPSL}
			\includegraphics[width=0.34\textwidth]{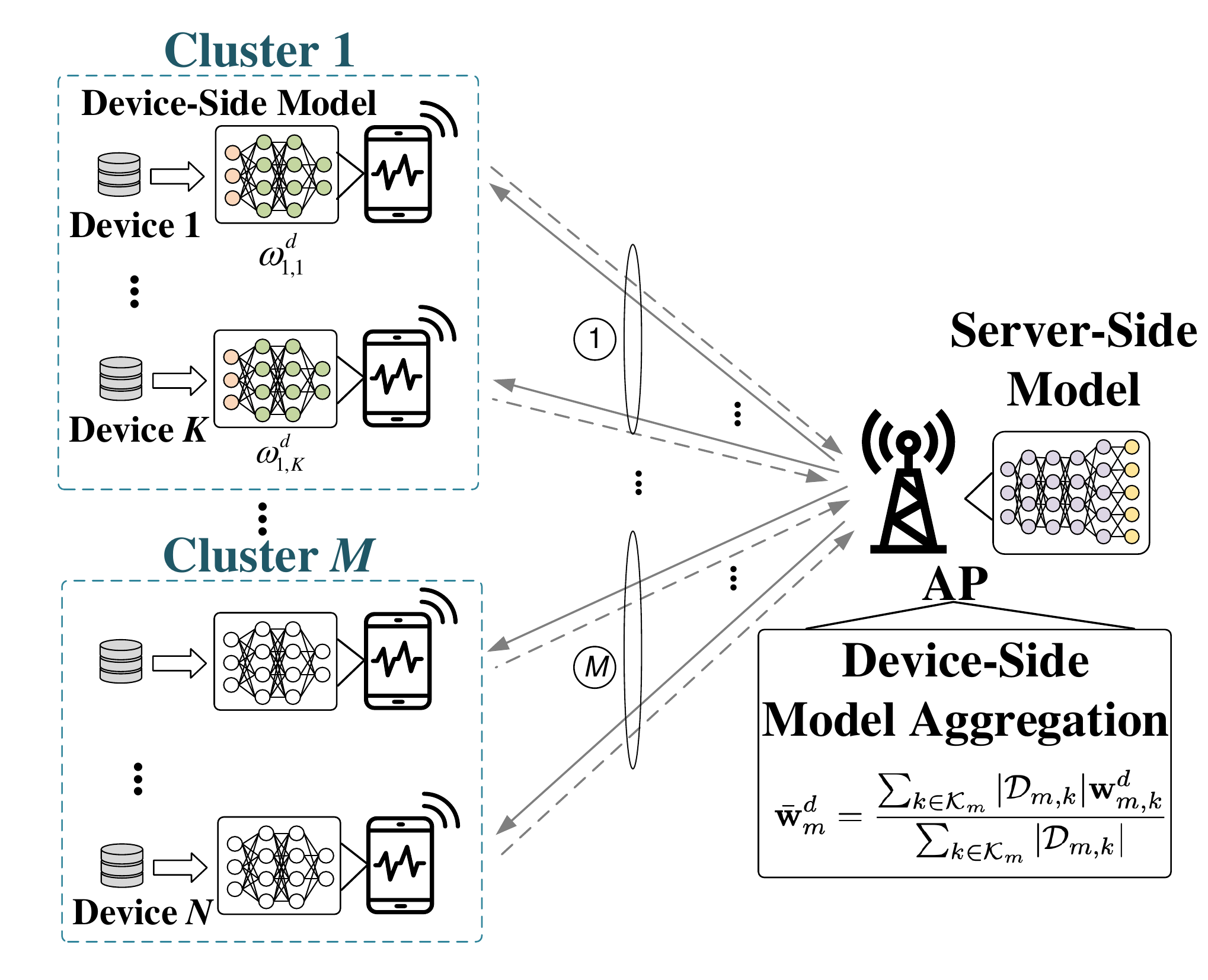}}
	\end{subfigure}
	\caption{(a) In the vanilla SL scheme, devices are trained sequentially; and (b) in the CPSL, devices are trained parallelly in each cluster while clusters are trained sequentially. }
	\vspace{-0.7cm}
\end{figure}

In this section, we present the low-latency CPSL scheme, as illustrated in Fig.~\ref{fig:CPSL}. The \emph{core idea} of the CPSL is to partition devices  into several clusters, parallelly train device-side models in each cluster and aggregate them, and then sequentially train the whole AI model across clusters. The detailed procedure of the CPSL is presented in Alg.~\ref{alg:CPSL}. 


\subsection{Initialization}
In the \emph{initialization} stage, the model parameter  is initialized randomly, and the optimal cut layer for minimizing training latency is selected using Alg.~\ref{algorithm:cut-layer} (to be discussed). After initialization, the CPSL operates in consecutive \emph{training rounds} until the optimal model parameter is identified. It is assumed that all the participating devices in one training round stay in AP's coverage area with static device channel conditions and computing capabilities. At each training round, $t\in\mathcal{T}= \{1, 2,..., T\}$, the following intra-cluster learning and inter-cluster learning stages are performed. 

\subsection{Intra-Cluster Learning Stage}
The intra-cluster learning stage is to facilitate parallel model training for devices within a cluster, consisting of the following steps.

\emph{Step 1 - Device clustering (Lines 3-4)}. In this step, the AP collects real-time information of participating device computing capabilities and channel conditions, and then partitions devices into multiple clusters. The device clustering decision making algorithm is detailed in Alg.~\ref{algorithm:scheduling} (to be discussed). Let $M$ denote the number of clusters, and $\mathcal{M}$ is the set of clusters.  The set of devices of  cluster $m$ is denoted by $\mathcal{K}_m, \forall m\in \mathcal{M}$. We have $\cup_{m=1}^{M} \mathcal{K}_m=\mathcal{N}$, and $\mathcal{K}_n \cup \mathcal{K}_m=\emptyset$ if $n \neq m$. The following steps are to facilitate parallel device-side model training in a cluster.



\begin{algorithm}[t]\label{alg:CPSL}
			\small
	\caption{Cluster-based parallel split learning (CPSL) scheme.}
	\LinesNumbered   

	\SetKwInOut{Input}{Input}
	\Input{$B$,  $\eta_d, \eta_e$, and, $K$; }
	\SetKwInOut{Output}{Output}
	\Output{$\mathbf{w}^\star$;}
	Initialize model parameter and determine the cut layer using Alg.~\ref{algorithm:cut-layer}\; 
	\For{training round $t=1,2,...,T$ }
	{
		AP collects real-time computing capabilities and channel conditions of all the devices\;
		AP partitionss devices into clusters using Alg.~\ref{algorithm:scheduling}\;
		\For{\text{cluster $m=1,2,...,M$} }
		{		
			AP broadcasts the latest device-side model to participating devices in cluster $m$\;
			\For{\text{local epoch $l=1,2,...,L$} }
			{
				\For{each device  \textbf{in parallel} }
				{
					Draw a mini-batch of data samples\;
					Execute device-side model and obtain smashed data via \eqref{equ:device-side FP}\;
					Transmit smashed data to the AP with allocated radio spectrum using  Alg.~\ref{algorithm:spectrum}\;
				}
				AP concatenates smashed data  and  executes the server-side model via \eqref{equ:server-side FP}\;
				AP updates the server-side model via \eqref{equ:BP_server-side}\;
				AP transmits smashed data's gradient to participating devices\;
				\For{each device \textbf{in parallel} }
				{
					Update the device-side model using \eqref{equ:BP_device-side}\;
				}
			}
			\For{each device  \textbf{in parallel} }
			{
				Upload the device-side model to the AP with allocated radio spectrum\;
			}
			AP aggregates device-side models into a new device-side model via \eqref{equ:Model aggregation}\;
		}	
	}
\end{algorithm}

\emph{Step 2 - Device-side model distribution (Line 6)}. In this step, the AP broadcasts the initial device-side model, denoted by $\mathbf{w}_m^d(t)$, to all the participating devices in cluster $m$. The AI model is trained for $L$ \emph{local epochs}, indexed by $l\in \mathcal{L} = \{1,2,..., L\}$. Let $\mathbf{w}^d_{m,k}(t, l) $ denote the device-side model parameters of device $k$ in cluster $m$ at epoch $l$ in training round $t$. In the first local epoch, we have
\begin{equation}
	\mathbf{w}^d_{m,k}(t,1)\leftarrow \mathbf{w}^d_{m}(t), \forall k\in \mathcal{K}_m.
\end{equation} 
All participating devices in a cluster share the same server-side model at each local epoch, denoted by $\mathbf{w}^e_{m}(t, l)$.
	
\emph{Step 3 - Model execution (Lines 8-13)}. 
This step is to execute the model to compute the predicted results based on drawn data samples, i.e., the FP process. Steps~2 and~3 are repeated for $L$ times. The whole AI model execution is split into two phases, including device-side model execution and server-side model execution.
\begin{itemize}
	\item \emph{Device-side model execution}: Firstly, each device randomly draws a \emph{mini-batch} of data samples, denoted by $\mathcal{B}_{m,k}(t,l)\subseteq \mathcal{D}_{m, k}$, from its local dataset. Here, $B=|\mathcal{B}_{m,k}(t,l)|$ is the mini-batch size, and $\mathcal{D}_{m,k}$ denotes the dataset possessed by device $k$. Let $\mathbf{Z}_{m,k} (t,l)\in \mathbb{R}^{B\times Q}, \forall k\in \mathcal{K}_m$ denote the aggregated input of the mini-batch of data samples in device $k$. Secondly, each device executes its respective device-side model with the drawn data samples, and obtains smashed data $\mathbf{S}_{m,k}(t,l) \in \mathbb{R}^{B\times P}$, i.e.,
	\begin{equation}\label{equ:device-side FP}
		\mathbf{S}_{m,k}(t,l)=f\left(\mathbf{Z}_{m,k}(t,l); \mathbf{w}^d_{m,k}(t,l)\right), \forall k\in \mathcal{K}_m, l\in \mathcal{L}
	\end{equation}
	where $f\left(\mathbf{z};\mathbf{w}\right)$ represents the mapping function between input $\mathbf{z}$ and output given model parameter $\mathbf{w}$. Here, $P$ is the dimension of smashed data for one data sample. Thirdly, each device transmits its smashed data to the AP with the allocated radio spectrum determined by Alg.~\ref{algorithm:spectrum} (to be discussed). 
	\item \emph{Server-side model execution}: The AP receives the smashed data from participating devices and then concatenates them into matrix $\mathbf{S}^{con}_m(t,l)=\left[\mathbf{S}_{m,1}(t,l); \mathbf{S}_{m,2}(t,l);...; \mathbf{S}_{m, K_m}(t,l)\right]\in \mathbb{R}^{K_mB\times P}$, which is fed into the server-side model $ \mathbf{w}^e_{m}(t,l)$. As such, the predicted result from the server-side model is given by
	\begin{equation}\label{equ:server-side FP}
		\hat{\mathbf{y}}(t,l)=f\left(\mathbf{S}_m^{con}(t,l); \mathbf{w}^e_{m}(t,l)\right)\in  \mathbb{R}^{K_mB\times 1}, \forall l\in \mathcal{L}.
	\end{equation}
	With \eqref{equ:device-side FP} and \eqref{equ:server-side FP}, the one-round FP process of the whole  model is completed.
\end{itemize}

	
	
\emph{Step 4 -  Model update (Lines 14-18)}. This step is to update the whole AI model by minimizing the loss function, which is the BP process. Similar to model execution, the model update includes device-side model update and server-side model update.
\begin{itemize}
	\item \emph{Server-side model update}: Given the predicted results and the corresponding ground-truth labels, the average gradient of the loss function  can be calculated and denoted by $\nabla l\left(\mathbf{w}\right)$. Then, the server-side model is updated by using the stochastic gradient descent  (SGD) method:
	\begin{equation}\label{equ:BP_server-side}
			\mathbf{w}_m^e(t,l+1)\leftarrow 	\mathbf{w}_m^e(t,l) - \eta_e {\nabla l\left(\mathbf{w}_m^e(t,l)\right)}, \forall l\in \mathcal{L}
	\end{equation}
	where $\eta_e $ is the learning rate for the server-side model update. The model parameters are updated layer-wise from the last layer to the cut layer according to the chain rule for gradient calculation. When the gradient calculation proceeds to the cut layer, the gradient of a minibatch of data samples, namely smashed data's gradient, is sent back to its corresponding device. 
	\item \emph{Device-side model update}: 	With the received smashed data's gradient, each device-side model  is  updated by using the SGD method:
	\begin{equation}\label{equ:BP_device-side}
		\begin{split}
				\mathbf{w}_{m,k}^d(t,l+1)\leftarrow  \mathbf{w}_{m,k}^d(t,l) - \eta_d   {\nabla l\left(\mathbf{w}_{m,k}^d(t,l)\right)} , \forall k \in \mathcal{K}_m, l\in \mathcal{L}
		\end{split}
	\end{equation}
	where $\eta_d$ is the learning rate for the device-side model update. With \eqref{equ:BP_server-side} and \eqref{equ:BP_device-side}, the one-round BP process is completed. 
\end{itemize}

 \emph{Step 5 - Model aggregation (Lines 20-23)}. This step is to aggregate the device-side models of participating devices in a cluster. After completing $L$ local epochs, the trained device-side models are uploaded to the AP and then aggregated via the FedAvg algorithm~\cite{bonawitz2019towards}. The aggregated device-side model is given by
	\begin{equation}\label{equ:Model aggregation}
		\bar{\mathbf{w}}_m^d(t)=\frac{\sum_{k\in \mathcal{K}_m}|\mathcal{D}_{m,k}|\mathbf{w}_{m,k}^d(t,L+1)}{\sum_{k\in \mathcal{K}_m}|\mathcal{D}_{m,k}|}.
	\end{equation}
In \eqref{equ:Model aggregation}, device-side models are aggregated via average based on the number of possessed data samples at each device.

\subsection{Inter-Cluster Learning}


This stage is to transfer the aggregated device-side model from one cluster to another cluster for continuing the training process, i.e.,
\begin{equation}\label{equ: Inter-cluster model training}
	\mathbf{w}_{m+1}^d(t)\leftarrow  \bar{\mathbf{w}}_{m}^d(t), \forall m=1,2,..., M-1.
\end{equation}
Then, the AP broadcasts the updated device-side model to devices in the next cluster. Each cluster conducts the intra-cluster learning stage until all clusters complete the training process. In this way, the inter-cluster learning stage is performed in a sequential manner across  clusters, which is similar to SL. 


\begin{remark}
	Different from the vanilla SL scheme that only operates in a sequential manner, the proposed CPSL operates in a “first-parallel-then-sequential” manner. Devices in each cluster are trained parallelly, while clusters are trained sequentially, thereby folding the entire training process and reducing the training latency. Extensive simulation results in Section~\ref{sec:Simulation Results} validate that the training latency can be significantly reduced.
\end{remark}




\section{Training Latency Analysis}\label{sec:Training Delay Analysis and Problem Formulation}
In this section, we present the decision variables in the proposed CPSL, based on which we analyze its training latency.
\subsection{Decision Variables in CPSL}
In the CPSL, the following decision variables should be determined.
\begin{itemize}
	\item \emph{Cut layer selection}: At the beginning of the entire training process, the cut layer selection decision, denoted by $v$, is determined beforehand based on historical data of participating devices. The decision is constrained by
	\begin{equation}\label{equ:cut_layer_decision}
		v \in \mathcal{V}
	\end{equation}
	where $\mathcal{V}=\{1,2, ..., V\}$ is the set of available cut layers in the considered AI model. Note that we consider a chain-topology DNN and $V$ is the number of DNN layers. A special case is that  cut layer $v=V$ means an empty server-side model. In other words, the CPSL scheme degrades to the FL scheme with $K_m$ devices.

	\item \emph{Device clustering}: At each training round, the device clustering decision is made based on the collected real-time  device channel conditions and computing capabilities, denoted by binary matrix $\mathbf{A}^t\in \mathbb{R}^{N \times M}, \forall t \in \mathcal{T}$. Each element is constrained by
	\begin{equation}\label{equ:device_clustering_decision}
		a_{n,m}^t\in \{0,1\}, \forall n\in \mathcal{N}, m\in 
		\mathcal{M},t\in \mathcal{T}
	\end{equation}
	where $a_{n,m}=1$ indicates that device $n$ is associated to cluster $m$, and $a_{n,m}=0$ otherwise.
	
	\item \emph{Radio spectrum allocation}: In each intra-cluster learning stage, we	consider the frequency-division multiple access for data transmission. Let $\left\{ \mathbf{x}_{1}^t, \mathbf{x}_{2}^t,..., \mathbf{x}_{M}^t \right\}$ denote the radio spectrum allocation decision where $\mathbf{x}_{m}^t\in \mathbb{Z}^{K_m \times 1}$ represents the decision in cluster $m$. Each element
	\begin{equation}\label{equ:spectrum-allocation_decion}
		x_{m,k}^t \in \mathbb{Z}^+, \forall k\in\mathcal{K}_m, m\in 
		\mathcal{M},t\in \mathcal{T} 
	\end{equation} 
	represents the number of subcarriers allocated to device $k$ in cluster $m$, where $\mathbb{Z}^+$ is the set of positive integers. Note that the number of  allocated subcarriers should not exceed the radio spectrum capacity, i.e.,
	\begin{equation}\label{equ:spectrum-allocation_decion_const}
		\sum_{k\in \mathcal{K}_m}x^t_{m,k}\leq C, \forall  m\in 
		\mathcal{M},t\in \mathcal{T}
	\end{equation} 
where $C$ represents the total number of subcarriers.
\end{itemize} 

\subsection{Training Latency}\label{sec:System Model and Problem Formulation}
The training latency of the CPSL is analyzed given the above decisions. The entire training process consists of multiple rounds, and each round consists of multiple  stages in each cluster. To characterize the overall training latency,  per-cluster training latency  is analyzed. For notation simplicity, we omit $t$ in this subsection.

\begin{figure}[t]
	\renewcommand{\figurename}{Fig.}
	\centering
	\includegraphics[width=0.45\textwidth]{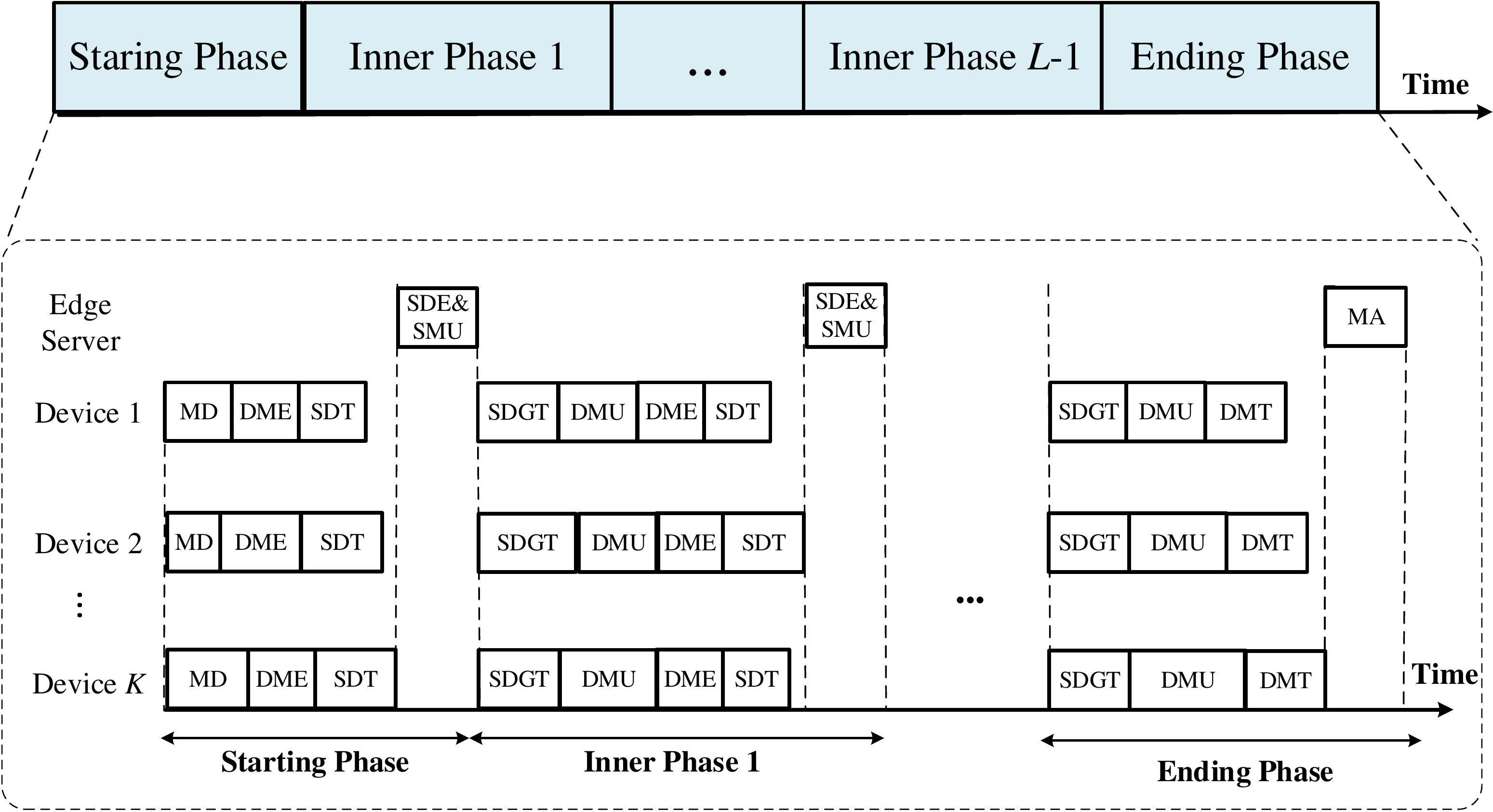}
	\caption{The procedure of the model training process in each cluster consists of a starting phase, multiple inner phases, and an ending phase.}
	\label{fig:Training_per_cluster}
				\vspace{-0.7cm}
\end{figure}

In the CPSL scheme, the AP waits for the update from all the participating devices, including two cases: (1) in each local epoch training, the AP waits for smashed data to perform server-side model execution; and (2) in each intra-cluster learning, the AP  waits for device-side models to perform model aggregation.  According to the AP's operation, the per-cluster training process can be divided into $L+1$ phases in a chronological manner, whose detailed structure is shown in Fig.~\ref{fig:Training_per_cluster}. The first one is a \emph{staring phase}, which spans from device-side model distribution to server-side model update in the first local epoch. The last one is an \emph{ending phase}, which spans from smashed data' gradient transmission in the last local epoch to device-side model aggregation. The remaining ones are $L-1$ identical \emph{inner phases}, each of which spans from smashed data's gradient transmission in the previous local epoch to server-side model update in the next local epoch. Note that if the number of local rounds equals 1, no inner phase exists. The detailed analysis is as follows.



\subsubsection{Starting Phase}
The latency of the starting phase includes the following four components.
\begin{itemize}
	\item \emph{Model distribution (MD) latency}: At the beginning of the CPSL scheme, the latest device-side model is broadcast to all the participating devices in the cluster using all  subcarriers. Let $\xi_d\left(v\right)$ denote the data size (in bits) of the device-side model, depending on cut layer $v$. The average downlink transmission rate of a subcarrier from the AP  to device $k$ is given by~\cite{liang2019spectrum} 
	\begin{equation}\label{equ:channel_rate}
		R_{k}^{DL}={\mathbb{E}}_{h_{k}}\left[{W}\log_2\left(1+\frac{P^{DL}\left|h_{k}\right|^2}{N_oW}\right)\right], \forall k \in \mathcal{K}_m
	\end{equation}
	where $W$, $P^{DL}$, $h_{k}$, and $N_o$ represent the subcarrier bandwidth,  AP's transmission power, channel gain, and thermal noise spectrum density, respectively.  Hence, the MD latency is given by
	\begin{equation}\label{equ:Device-side model distribution latency}
		{\tau_{b,k}=\frac{\xi_d\left(v\right)}{C R_{k}^{DL}}, \forall k\in \mathcal{K}_m}.
	\end{equation}
	
	\item \emph{Device-side model execution (DME) latency}:  The device-side model execution refers to the device-side model's FP process.  Let {$\gamma_d^F(v)$} denote the computation workload (in FLOPs) of device-side model's FP process for processing one data sample~\cite{zhang2018shufflenet, zeng2021energy}. The device-side model execution needs to process a mini-batch of data samples, and the overall computation workload is   {$B \gamma_d^F(v)$}.  The DME latency is given by
	\begin{equation}\label{equ: FP_device_side}
		{	\tau_{d,k}=\frac{B\gamma_d^F(v)}{f_k\kappa }, \forall k\in \mathcal{K}_m}
	\end{equation}
	where $f_{k}$ denotes the central processing unit (CPU) capability of device $k$, and $\kappa$ denotes the computing intensity.\footnote{The value of $\kappa$  represents the number of FLOPs can be completed in one CPU cycle, which is determined by the processor architecture. 	Note that the above latency analysis that can be readily extended to the case using graphics processing units (GPUs), in which GPUs take a different value of $\kappa$~\cite{zeng2021energy, zhang2018shufflenet}.}	
	
	\item 	\emph{Smashed data transmission (SDT) latency}: Each device transmits the smashed data to the AP using the allocated radio spectrum. Let $\xi_s(v)$ denote the smashed data size with respect to one data sample, also depending on cut layer $v$. For a minibatch of $B$ data samples,  the transmitted smashed data size in bits is represented by $B\xi_s(v)$.	For device $k$, the number of allocated subcarriers is given by $\mathbf{x}_{m, k}$. Similar to \eqref{equ:channel_rate}, the average uplink transmission rate of a subcarrier for device $k$ is given by $R_{k}^{UL}={\mathbb{E}}_{h_{k}}\left[{W}\log_2\left(1+{P^{UL}\left|h_{k}\right|^2}/{N_oW}\right)\right]$, where  $P^{UL}$ represents transmission power of a device.\footnote{Note that  we consider the time division duplex in the network,  such that uplink and downlink channel conditions can be assumed to be identical.} Hence, the SDT latency is given by
	\begin{equation}\label{equ:Smashed data transmission latency}
		{\tau_{s,k}=\frac{B\xi_s(v) }{x_{m,k}R_{k}^{UL}}, \forall k\in \mathcal{K}_m}.
	\end{equation}

	\item \emph{Server-side model execution (SME) and server-side model update (SMU) latency}: 
	The latency component includes two parts: (1) the SME latency represents the time taken for performing the server-side model's FP process. Let {$\gamma_s^F(v)$} denote the computation workload of the server-side model's FP process for processing one data sample.	Since all the smashed data are fed for training the server-side model, the number of the concatenated smashed data samples is $K_m B$, and the overall computation workload is $K_m B\gamma_s^F(v)$. Similar to \eqref{equ: FP_device_side}, the SME latency is given by {${K_m B \gamma_s^F (v)}/{f_s\kappa}$}, where $f_s$  denotes the CPU capability of the edge server; and (2) the second part is the time taken for performing the BP process of the server-side model. Let $\gamma_s^B(v)$ represent the computation workload of the server-side model's BP process for one data sample. Similarly, the SMU latency is given by {${K_m B \gamma_s^B(v)}/{f_s\kappa}$.}  Taking the two parts into account, the overall latency is given by
	\begin{equation}\label{equ: server-side operation}
		{\tau_{e}=\frac{K_m B \left(\gamma_s^F(v)+\gamma_s^B(v)\right)}{f_s\kappa}.}
	\end{equation}

\end{itemize}


Taking the latency components in \eqref{equ:Device-side model distribution latency}, \eqref{equ: FP_device_side},  \eqref{equ:Smashed data transmission latency}, and \eqref{equ: server-side operation} into account, the overall latency of the starting phase is given by
\begin{equation}\label{equ: initial_phase}
		d_{m}^S=	\max_{k\in \mathcal{K}_m} \{\tau_{b,k}+ 	\tau_{d,k}+\tau_{s,k}\}
		+	\tau_{e}, \forall m\in \mathcal{M}
\end{equation}
where the first term, $\max_{k\in \mathcal{K}_m}\{\tau_{b,k}+\tau_{d,k}+\tau_{s,k}\}$, is to account that all the smashed data should be received before server-side model execution.


\subsubsection{Inner Phase} 
The latency of each inner phase includes five  components from smashed data's gradient transmission (SDGT), device-side model update (DMU),  DME,  SDT,  SME, and SMU. The last four latency components have been analyzed above, and we analyze the first two components.
\begin{itemize}
	\item  \emph{SDGT latency}: After SME and SMU are performed, smashed data's gradient is sent back to each device using the allocated radio spectrum. Let {$\xi_g(v)$} denote the data size of smashed data's gradient.  Similar to \eqref{equ:Smashed data transmission latency}, the   latency is given by
	\begin{equation}\label{equ:Smashed data's gradient transmission latency}
		{\tau_{g,k}=\frac{\xi_g(v) }{x_{m,k}R_{k}^{DL}}, \forall k\in \mathcal{K}_m.}
	\end{equation}
	\item \emph{DMU latency}: The device-side model update refers to the BP process updating device-side model parameters. Let $\gamma_d^B(v)$ represent the computation workload of the device-side model's BP process  for one data sample.   Similar to \eqref{equ: server-side operation}, we have
	\begin{equation}\label{equ:Device-side model update latency}
		{\tau_{u,k}=\frac{B\gamma_d^B(v)}{f_k\kappa }, \forall k\in \mathcal{K}_m.}
	\end{equation}
	The cut layer affects the computation workload distribution between the device and the edge server. The total computation workload in the BP process is given by $\gamma^b= \gamma_d^B (v)+\gamma_s^B (v)$.\footnote{Regarding the FP process, $\gamma^F= \gamma_d^F (v)+\gamma_s^F (v)$, where $\gamma^F$ is the computation workload of the whole FP process.}  A shallow cut layer means heavy computation workloads on the edge server, while a deep cut layer means heavy computation workloads on the device.
\end{itemize}

Similar to \eqref{equ: initial_phase}, taking all latency components into account, the overall latency in each inner phase is given by
\begin{equation}\label{equ:inner_phase}
		d_{ m}^I =\max_{k\in \mathcal{K}_m} \{\tau_{g,k}+\tau_{u,k}+ \tau_{d,k}+\tau_{s,k}\}+	\tau_{e}, \forall m\in \mathcal{M}.
\end{equation}

\subsubsection{Ending Phase}
The ending phase includes four latency components from SDGT, DMU, device-side model transmission (DMT), and model aggregation (MA). The  first two components are analyzed above, and we analyze the rest two components. Regarding the DMT latency, each device transmits its device-side model to the AP using the allocated radio spectrum, and the corresponding latency is given by 
\begin{equation}
	{\tau_{t,k}=\frac{\xi_d(v)}{x_{m,k}R_{k}^{UL}}, \forall k\in \mathcal{K}_m.}
\end{equation} 
The MA latency is negligible since  aggregating models using the FedAvg algorithm incurs a relatively low computational complexity. Taking all the latency components into account, the overall latency in the ending phase is given by
\begin{equation}\label{equ:final_phase}
	{d_{m}^E= \max_{k\in \mathcal{K}_m} \{\tau_{g,k}+\tau_{u,k}+ \tau_{t,k}\},} \forall m\in \mathcal{M}
\end{equation}
where the maximization operation is to account that MA has to wait for the straggler device.


\subsubsection{Overall Training Latency}
With the results of all the phases in \eqref{equ: initial_phase}, \eqref{equ:inner_phase}, and \eqref{equ:final_phase}, the per-cluster training latency is given by $D_m\left(v,\mathbf{A}^t, \mathbf{x}^t_m \right)=d_{m}^S+	\left(L-1\right)d_{ m}^I +d_{m}^E$.  The overall training latency of the  CPSL scheme in one training round with $M$ clusters is given by
\begin{equation}
	{D}^t\left(v, \mathbf{A}^t, \{\mathbf{x}^t_m\}_{ m\in \mathcal{M} } \right)= \sum_{m\in \mathcal{M}} D_m\left(v, \mathbf{A}^t, \mathbf{x}^t_m \right),
\end{equation}
which depends on device clustering decision $\mathbf{A}^t$, radio spectrum allocation decision $\{\mathbf{x}^t_m\}_{ m\in \mathcal{M} }$, and cut layer selection decision $v$. Considering all training rounds, the overall latency is
\begin{equation}
\bar{D}=\sum_{t\in \mathcal{T}} {D}^t\left(\mathbf{A}^t,\{\mathbf{x}^t_m\}_{m\in \mathcal{M}}, v\right).	
\end{equation}
In the following, these decisions are optimized to minimize the training latency of the CPSL.
\begin{remark}
	The cut layer selection decision determines not only communication overhead since the data sizes of the device-side model, smashed data, and smashed data's gradient depend on the cut layer, but also computation workload distribution between the device and the edge server. As such, the cut layer selection plays an important role in optimizing the training latency. 
\end{remark}


\section{Resource Management Problem Formulation and Decomposition} \label{sec: problem formulation}
\subsection{Problem Formulation}
Since device computing capabilities and channel conditions vary temporally, minimizing the long-term overall training latency is paramount. The proposed CPSL scheme requires jointly making cut layer selection, device clustering, and radio spectrum allocation decisions. To this end, we formulate the resource management problem to minimize the overall training latency:

\begin{subequations}\label{Problem 0}
	\begin{align}
		{\mathcal{P}:}	 \underset{\begin{subarray}{c} v, \{\mathbf{A}^t\}_{t\in \mathcal{T}},\\ \{\mathbf{x}_m^t\}_{    \begin{subarray}{c}
					m\in \mathcal{M}\\ t\in \mathcal{T}
		\end{subarray}  }	\end{subarray}}{\text{min}}\;\;\;
		 & 	\sum_{t\in \mathcal{T}} {D}^t\left(v, \mathbf{A}^t,\{\mathbf{x}^t_m\}_{m\in \mathcal{M}}\right)	 \label{Constraint_P0_0}\\  
		 \text{s.t.}\;\;\;
		& \sum_{n \in \mathcal{N}} a_{n,m}^t=K_m, \forall m\in 
		\mathcal{M}, t\in \mathcal{T}, \label{Constraint_P0_2}\\
		& \eqref{equ:cut_layer_decision}, \eqref{equ:device_clustering_decision}, \eqref{equ:spectrum-allocation_decion}, \text{and } \eqref{equ:spectrum-allocation_decion_const}.\nonumber
	\end{align}
\end{subequations} 
Constraint \eqref{Constraint_P0_2} guarantees the number of devices in each cluster satisfies cluster capacity limit, and constraints $\eqref{equ:cut_layer_decision}$, $\eqref{equ:device_clustering_decision}$, $\eqref{equ:spectrum-allocation_decion}$, and $\eqref{equ:spectrum-allocation_decion_const}$ guarantee feasible decision variables. 

Problem  $\mathcal{P}$  is a \emph{stochastic mix-timescale} optimization problem. The problem is “stochastic” because the decisions are determined in presence of  temporal dynamics of device computing capabilities and channel conditions during the training process. The problem is “mix-timescale” because the decisions are made in different timescales. The cut layer selection is determined for the entire training process (i.e., in a large timescale), while the device clustering and radio spectrum allocation  are determined for each training round (i.e., in a small timescale). The device clustering and radio spectrum allocation decisions are coupled with each other, which further complicates the problem.


\subsection{Problem Decomposition}
To solve problem $\mathcal{P}$, we first decompose it into the following  subproblems in two timescales by exploiting the timescale separation of the decision variables.

\textbf{Subproblem 1: Large-timescale cut layer selection subproblem}. The optimal cut layer is selected for the entire training process to minimize the overall training latency, i.e., 
	\begin{subequations}\label{Subproblem 1}
		\begin{align}
			{\mathcal{P}_L:}\;\;	 \underset{v}{\text{min}}\;\;
			 & 		\sum_{t\in \mathcal{T}} {D}^t\left(v, \mathbf{A}^t,\{\mathbf{x}^t_m\}_{m\in \mathcal{M}}\right) \nonumber\\  
			 \text{s.t.}\;\;
			 & \eqref{equ:cut_layer_decision}. \nonumber
		\end{align}
	\end{subequations} 
The above objective function is non-convex, because not only the data sizes of smashed data, smashed data's gradient, and the device-side model, but also the computation workloads of the device-side model's FP and BP processes are arbitrary functions with respect to the cut layer.

 \textbf{Subproblem 2: Small-timescale device clustering and radio spectrum allocation subproblem.} At each training round~$t$, the device clustering  and  radio spectrum allocation decisions are jointly optimized to minimize the one-round training latency:
	\begin{subequations}\label{Problem joint}
		\begin{align}
			{\mathcal{P}_S:}	 \underset{\mathbf{A}^t, \{\mathbf{x}^t_m\}_{m\in 
					\mathcal{M}}}{\text{min}}\;\;
			& 		D^t\left(v, \mathbf{A}^t, \{\mathbf{x}^t_m\}_{m\in 
				\mathcal{M}} \right) \nonumber\\  
			  \text{s.t.}\;\;\;\;\;\;\;\;\;
			  & \eqref{equ:device_clustering_decision}, \eqref{equ:spectrum-allocation_decion}, \eqref{equ:spectrum-allocation_decion_const}, \text{and } \eqref{Constraint_P0_2}.\nonumber
		\end{align}
	\end{subequations} 
	In the subproblem, the optimization variables are integer. Hence, the problem is a combinatorial optimization problem, which is NP-hard (one of Karp’s 21 NP-complete problems~\cite{karp1972reducibility}). Moreover, according to the definitions of latency components in \eqref{equ:inner_phase} and \eqref{equ:final_phase}, the objective function is to minimize the maximum latency among all the participating devices in different stages, which is non-convex. As such, a low-complexity algorithm is desired.



	The \emph{relationship} between the above two subproblems is as follows. Given optimal cut layer $v^\star$ by solving Subproblem~$\mathcal{P}_L$, the optimal device clustering decision, $(\mathbf{A}^t)^\star$, and radio spectrum allocation decision, $ \{\mathbf{x}^t_m\}_{m\in 	\mathcal{M}}^\star$,  can be obtained via solving Subproblem~$\mathcal{P}_S$ based on real-time network information at each training round. As such, $\left\{ v^\star, (\mathbf{A}^t)^\star, \{\mathbf{x}^t_m\}_{m\in 
		\mathcal{M}}^\star\right\}$ is the optimal solution for problem~$\mathcal{P}$.

\section{Two-Timescale Resource Management Algorithm}\label{sec:resource Management Algorithm}
In this section, a two-timescale resource management algorithm is proposed to jointly solve problem~$\mathcal{P}$,  consisting of an SAA-based cut layer selection algorithm and a Gibbs sampling-based joint device clustering and radio spectrum allocation algorithm.



\subsection{Large Timescale: Cut Layer Selection Algorithm} 
In this subsection, we present the SAA-based algorithm to determine the optimal cut layer, consisting of the following steps.

Firstly, the objective function in problem $\mathcal{P}_L$ can be approximated by 
\begin{equation}\label{equ:function_approximation}
	\sum_{t\in \mathcal{T}} {D}^t\left(v, \mathbf{A}^t,\{\mathbf{x}^t_m\}_{m\in \mathcal{M}}\right) \approx T \mathbb{E}_{\mathbf{f}, \mathbf{h}}\left[{D}\left(v, \mathbf{A},\{\mathbf{x}_m\}_{m\in \mathcal{M}}\right)\right].
\end{equation}
In \eqref{equ:function_approximation}, $\mathbb{E}_{\mathbf{f}, \mathbf{h}}\left[{D}\left(v, \mathbf{A},\{\mathbf{x}_m\}_{m\in \mathcal{M}}\right)\right]$ represents the average per-round training latency, where $\mathbf{f}=[f_1, f_2,..., f_N]$ and $\mathbf{h}=[h_1, h_2,..., h_N]$ denote random variables of device computing capabilities and channel conditions, respectively. We assume that $f_n$ follows a Gaussian distribution with mean $\mu_{n,f}$ and variance $\sigma_f^2$, i.e., $f_n \sim {N}\left(\mu_{n,f}, \sigma_f^2 \right), \forall n \in \mathcal{N}$, due to time-varying device computation workloads. Similarly, we assume  $h_n \sim {N}\left(\mu_{n,h}, \sigma_h^2 \right)$ due to shadowing effect in wireless channels. It is worth noting that the proposed algorithm can  be applied to arbitrary distribution settings. The number of training rounds, $T$, depends on many factors, such as model structure and data distribution, which are independent of the decision variables. In this way, we aim to minimize the average per-round training latency.

Secondly, we leverage the SAA method~\cite{tang2019service, sun2019qoe} to  approximate the average per-round training latency. The core idea of the SAA is to approximate the expectation of a random variable by its sample average. Specifically, several samples are drawn from the historical data of device computing capabilities and channel conditions to approximately compute the average per-round training latency. Let $J$ denote the number of samples. For sample $j$, given the device computing capabilities and channel conditions, the corresponding device clustering and radio spectrum allocation decisions, $\mathbf{A}^j$ and $\{\mathbf{x}^j_m\}_{m\in \mathcal{M}}$, can be obtained using Alg.~\ref{algorithm:scheduling}. As such, the sample-wise training latency is represented by $D\left( v, \mathbf{A}^j, \{\mathbf{x}^j_m\}_{m\in \mathcal{M}} \right)$, and the average per-round training latency can be approximated by
\begin{equation}\label{equ: approximation}
	\mathbb{E}_{\mathbf{f}, \mathbf{h}}\left[{D}\left(v, \mathbf{A},\{\mathbf{x}_m\}_{m\in \mathcal{M}}\right)\right]\approx \frac{1}{J} \sum_{j=1}^{J} D\left( v, \mathbf{A}^j, \{\mathbf{x}^j_m\}_{m\in \mathcal{M}} \right).
\end{equation}
For a large value of $J$, such approximation is valid. Thus, the problem of finding the optimal cut layer  can be converted into: 
\begin{subequations}\label{Problem C}
	\begin{align}
		{\mathcal{P}^C:}\;\;	 \underset{v}{\text{min}}\;\;
		 & 		\frac{1}{J} \sum_{j=1}^{J} D\left(\mathbf{A}^j, \{\mathbf{x}^j_m\}_{m\in \mathcal{M}}, v \right) \nonumber\\  
		 \text{s.t.}\;\;
		 & \eqref{equ:cut_layer_decision}.\nonumber
	\end{align}
\end{subequations}



Thirdly, the  problem can be solved via an exhaustive search method for a finite number of DNN layers. The detailed procedure of the proposed SAA-based algorithm is presented in Alg.~\ref{algorithm:cut-layer}. Specifically, given a cut layer, we can leverage the device clustering and radio spectrum allocation algorithm to calculate the average per-round training latency for all $J$ samples. After examining all the possible cut layers, optimal cut layer $v^\star$ can be determined.  The exhaustive search based algorithm can be conducted by the AP equipped with a high-end edge server in an offline manner, such that the computational complexity is affordable.


\begin{algorithm}[t]	\label{algorithm:cut-layer}
		\small
	\SetAlgoLined
	\LinesNumbered   
	\SetKwInOut{Input}{Input}
	Randomly draw $J$ samples of device computing capabilities and channel conditions from historical data\;
	\For{each cut layer $v \in \mathcal{V}$}
	{
		Calculate the expected training latency  $\Delta\left(v\right)= 	\frac{1}{J} \sum_{j=1}^{J} D^t\left(v, \mathbf{A}^j, \{\mathbf{x}^j_m\}_{m\in \mathcal{M}} \right)$ based on $J$ data samples using Alg.~\ref{algorithm:scheduling}\;
	}
	$v^\star = \arg\min_{v\in \mathcal{V}} \{\Delta\left(v\right)\} $.
	\caption{SAA-based cut layer selection algorithm.}
\end{algorithm}

\subsection{Small Timescale: Joint Device Clustering and Radio Spectrum Allocation Algorithm}
In each training round, device clustering and radio spectrum allocation decisions are jointly determined to minimize instantaneous one-round training latency based on real-time device computing capabilities and channel conditions.  For notation simplicity, we  omit $t$ in $\mathbf{A}^t$ and $ \{\mathbf{x}^t_m\}_{m\in \mathcal{M}}$ in this subsection. 

The device clustering and radio spectrum allocation decisions exhibit different properties. Given the device clustering decision, radio spectrum allocation decisions in each cluster are made independently. Moreover, the optimal spectrum allocation decision can be easily obtained via a greedy-based subroutine. Leveraging such property, we can decouple problem  $\mathcal{P}_S$  into a device clustering subproblem in the outer layer and multiple radio spectrum allocation subproblems in the inner layer, and propose a joint solution for them.



\begin{algorithm}[t]	\label{algorithm:spectrum}
	\small
	\SetAlgoLined
	\LinesNumbered   
	\SetKwInOut{Input}{Input}
	Initialization: ${x}_{m, k} =1, \forall k\in \mathcal{K}_m$\;
	\For{iteration $=1, 2, ..., C-K_m$}
	{
		$\Omega = D_{m}\left(v^\star, \mathbf{A}, \mathbf{x}_m \right)$\;
		\For{$k= 1,2,..., {K}_m$}
		{
			$\hat{x}_{m,k}= x_{m,k}+1 $\;
			$\hat{\mathbf{x}}_m=\{{x}_{m, 1}, {x}_{m, 2},..., \hat{x}_{m,k},..., {x}_{m, K_m}\}$\;
			$\Omega_k = D_{m}\left(v^\star, \mathbf{A }, \hat{\mathbf{x}}_m \right)$\;
		}
		$k^\star = \arg\max_{k\in \mathcal{K}_m} \{\Omega- \Omega_k\} $\;
		${x}_{m,k^\star} = x_{m,k^\star}+1$\;	
	}
	\caption{Greedy-based radio spectrum  allocation subroutine.}
\end{algorithm}

\subsubsection{Radio Spectrum Allocation Subproblem}
The proposed CPSL scheme sequentially trains clusters, such that the training latency is accumulated across clusters.  In addition, radio spectrum allocation decisions are independent among clusters. As such, optimizing the per-round training latency problem can be converted to individually optimizing the training latency in each cluster. The radio spectrum (subcarrier) allocation subproblem for each cluster in the inner layer  can be formulated~as:
\begin{subequations}\label{Problem 1}
	\begin{align}
		{\mathcal{P}^{S}:}\;\;	 \underset{\mathbf{x}_m}{\text{min}}\;\;
		 & D_{m}\left(v^\star, \mathbf{A}, \mathbf{x}_m \right)  \\
		 \text{s.t.}\;\;
		 & \eqref{equ:spectrum-allocation_decion}, \text{and } \eqref{equ:spectrum-allocation_decion_const}. \nonumber 
	\end{align}
\end{subequations} 
Since decision variable $\mathbf{x}_m$ is integer, problem $\mathcal{P}^{S}$ is an integer optimization problem with a non-convex objective function, which cannot be solved via existing convex optimization methods.


To solve the problem efficiently, we propose a greedy-based radio spectrum allocation subroutine by leveraging the \emph{diminishing gain} property of the problem. The diminishing gain property~\cite{wu2021learning} means that, in the subcarrier allocation problem, the gain of reducing latency decreases with the number of allocated subcarriers. Hence, the available radio spectrum should be allocated to the device that can achieve the maximum gain.  Specifically, the radio spectrum allocation decision is first initialized by allocating each device with one subcarrier. Then, an additional subcarrier is allocated to the device that can reduce the per-cluster training latency most until all the subcarriers have been allocated, as detailed in Alg.~\ref{algorithm:spectrum}.

\subsubsection{Device Clustering Subproblem}
The device clustering subproblem in the outer layer is to determine the optimal device clustering decision, which can be formulated as follows:
\begin{subequations}\label{Problem outer}
	\begin{align}
		{\mathcal{P}^D:}\;\;	 \underset{\mathbf{A} }{\text{min}}\;\;
		 & 		D\left(v^\star, \mathbf{A}, \{\mathbf{x}_m\}_{m\in 
			\mathcal{M}}   \right)\\
		 \text{s.t.}\;\;
		& \eqref{equ:device_clustering_decision}, \text{and } \eqref{Constraint_P0_2}. \nonumber 
	\end{align}
\end{subequations}


The  device clustering subproblem is a binary optimization problem with the cluster capacity constraint. To solve the problem efficiently, we propose a device clustering algorithm based on the Gibbs sampling method, which can determine the optimal device clustering in an iterative manner~\cite{robert2013monte, xu2018joint}. Let $G$ denote the number of iterations until convergence.  The radio spectrum allocation subroutine is embedded into the device clustering algorithm, such that the device clustering and radio spectrum allocation decisions are jointly determined. 

\subsubsection{Joint Device Clustering and Radio Spectrum Allocation Algorithm}
The joint algorithm is presented in Alg.~\ref{algorithm:scheduling}, which consists of the following two steps.

\begin{algorithm}[t]	\label{algorithm:scheduling}

		\small
	\SetAlgoLined
	\LinesNumbered   
	\SetKwInOut{Input}{Input}
	\Input{Real-time device computing capabilities $\mathbf{f}$ and channel conditions $\mathbf{h}$;}
	\SetKwInOut{Output}{Output}
	\Output{$\mathbf{A}$ and $\{\mathbf{x}_m\}_{m\in 
			\mathcal{M}}$;}
	{Initialization}: Randomly take a feasible device clustering decision $\mathbf{A}$, and obtain objective function value ${\Theta}= D\left(v^\star, \mathbf{A},\{\mathbf{x}_m\}_{m\in 
		\mathcal{M}}  \right)$\;
	\For{iteration $=1, 2,..., G$}{
		{$\rhd$ New decision generation}\\
		Randomly choose device $n \in \mathcal{K}_m$ and device $ n'\in \mathcal{K}_{m'}$ in two random clusters $m$ and $ m'$\; 
		Swap device association via $\hat{a}_{n,m} = a_{n',m'}$, $\hat{a}_{n',m'} = a_{n,m}$\;
		Obtain a new device clustering decision $\hat{\mathbf{A}}=\{a_{1,1}, a_{1,2},..., \hat{a}_{n,m},..., \hat{a}_{n',m'},...,  a_{N, M}\}$\;
		\For{each cluster $m\in \mathcal{M}$}{
			Obtain the optimal radio spectrum allocation decision  $\{\hat{\mathbf{x}}_m\}$ in each cluster using Alg.~\ref{algorithm:spectrum}\;
		}	
		{$\rhd$  Decision update}\\
		Obtain  objective function value by $\hat{\Theta}= D\left(v^\star , \hat{\mathbf{A}}, \{\hat{\mathbf{x}}_m\}_{m\in 
			\mathcal{M}} \right)$ given $\hat{\mathbf{A}}$ and $\{\hat{\mathbf{x}}_m\}_{m\in \mathcal{M}}$\;
		Set $\epsilon$ according to \eqref{equ: eplsion update}\; 
		Set $\{{\mathbf{A}}, {\Theta}\} =\{ \hat{\mathbf{A}}, \hat{\Theta}\}$ with probability $\epsilon$; otherwise, keep $\mathbf{A}$ and ${\Theta}$ unchanged.  
	}
	\caption{Joint device clustering and radio spectrum allocation algorithm.}
\end{algorithm}

\begin{itemize}
	\item \emph{New decision generation}: Two random devices are selected from two random clusters, denoted by device $m$ in cluster $n$ and device $m'$ in cluster $n'$. The corresponding device clustering decisions are swapped such that the capacity constraint in \eqref{equ:device_clustering_decision} is not violated, i.e., $\hat{a}_{n,m} \leftarrow a_{n',m'}$, $\hat{a}_{n',m'} \leftarrow a_{n,m}$. As such, we obtain a new device clustering decision $\hat{\mathbf{A}}=\{a_{1,1}, a_{1,2},..., \hat{a}_{n,m},..., \hat{a}_{n',m'},...,  a_{N,N_N}\}$. Given the new device clustering decision, the optimal radio spectrum allocation decisions for each cluster  $\{\hat{\mathbf{x}}_m\}_{m\in \mathcal{M}}$ can be solved using Alg.~\ref{algorithm:spectrum}. 
	
	\item \emph{Decision update}:  Given the joint device clustering and radio spectrum allocation decisions, the corresponding objective function value can be  obtained via $\hat{\Theta}\leftarrow D\left( v^\star, \mathbf{A}, \{\hat{\mathbf{x}}_m\}_{m\in 
		\mathcal{M}}  \right)$. Determine the probability of decision update via
	\begin{equation}\label{equ: eplsion update}
		\epsilon=\frac{1}{{1+e^{{\left(\hat{\Theta}-\Theta\right)}/{\delta}}}}.
	\end{equation} 
	In \eqref{equ: eplsion update}, $\delta$ is the smooth factor to control the tendency of new decision exploration. A larger value of $\delta$ tends to explore new decisions with a higher probability. Then, with probability $\epsilon$, the updated device clustering decision is taken; otherwise, the device clustering decision remains the same. 
\end{itemize}

When $\delta$ approaches 0, the algorithm converges to the global optima with probability~1~\cite{xu2018joint}.

\section{Simulation Results}\label{sec:Simulation Results}
\label{sec:erformance Evaluation of Parallel}


\subsection{Simulation Setup}
We conduct extensive simulations to evaluate the performance of the proposed CPSL scheme and the resource management algorithm. Below we introduce the key components of the simulation. The main simulation parameters are listed in Table~\ref{Simulation parameters}.

\begin{table}[t]
	\scriptsize
	\vspace{-0.5cm}
	\centering
	\caption{Simulation Parameters.}
	\label{Simulation parameters}
	\vspace{-0.5cm}
	\begin{tabular}{|c|c|c|c|c|c|c|c|}
		\hline
		\hline
		\textbf{Parameter} & \textbf{Value} &\textbf{Parameter} & \textbf{Value} & \textbf{Parameter} & \textbf{Value} &\textbf{Parameter} & \textbf{Value}\\
		\hline
		$f_s$& 100\;GHz & $f_n$& 0.5\;GHz &
		$\xi_s$ & 18 KB & $\xi_g$ & 36.1 KB \\\hline
		$\xi_d$& 0.67\;MB & $\tau$  & 1 &
		$B$ & 16 & $N$ & 30\\\hline
		$D_n$ & 180 &	$W$ & 1 MHz&
		$C$  & 30  &$\gamma_d^B$& 5.6\;MFlops \\\hline
		$\gamma_d^F$& 5.6\;MFlops &$\gamma_s^B$& 86.01\;MFlops &
		$\gamma_s^F$& 86.01\;MFlops  & $K$ & 1\\\hline
		$\delta$ &0.0001 & $G$&1,000&
		$Q$ & 100 &$N_m$ &5 \\ \hline
		$\eta_d$ & 0.05 &$\eta_e$ &0.25 & & & & \\
		\hline
	\end{tabular}
\end{table}	

\begin{table}[t]
	\scriptsize
	\vspace{-0.5cm}
	\centering
	\caption{AI Model Structure.}
	\label{DNN model parameters}
	\vspace{-0.5cm}
	\begin{tabular}{|c|c|c|c|c|c|c|c|}
		\hline
		\hline
		\textbf{Index} &	\textbf{Layer Name} & \textbf{NN Units} &\textbf{Activation} &	\textbf{Index} &	\textbf{Layer Name} & \textbf{NN Units} &\textbf{Activation}  \\
		\hline
		1&	CONV1& 32, $3 \times 3$   & ReLU & 	7&	CONV5& 128, $3 \times 3$   & ReLU \\\hline
		2&	CONV2& 32, $3 \times 3$   & ReLU & 	8&	CONV6& 128, $3 \times 3$   & ReLU \\\hline
		3&	POOL1& $2 \times 2$ & None &	9&	POOL3& $2 \times 2$ & None \\\hline
		4&	CONV3& 64, $3 \times 3$   & ReLU &		10&	FC1& 382& ReLU \\\hline
		5&	CONV4& 64, $3 \times 3$   & ReLU & 11&	FC2& 192& ReLU \\\hline
		6&	POOL2& $2 \times 2$ & None &	12&	FC3& 10& Softmax \\\hline
	\end{tabular}
	\vspace{-1cm}
\end{table}

The computing capability of the edge server is set to 100$\times 10^9$\;cycles/s. The number of devices is set to 30, and the radio spectrum bandwidth is set to 30\;MHz, unless otherwise specified. The subcarrier bandwidth  is set to 1\;MHz. Two image classification datasets are used in the simulation: (1) \emph{MNIST dataset}, where each data sample is an image associated with a label from ten classes of handwritten digits from ``0'' to ``9''~\cite{lecun1998gradient}; and (2) \emph{Fashion-MNIST dataset}, where each data sample is also an image with a label associated with ten clothing classes, such as ``Shirt'' and ``Trouser''~\cite{xiao2017fashion}. Both datasets consist of a {training dataset} with 50,000 data samples for model training and a {test dataset} with 10,000 data samples for performance evaluation. In addition, data distribution at devices is non-IID, which widely exists in practical systems. We assume that each device has only three classes of data samples, and these three classes are randomly selected among ten classes. Each device possesses 180 data samples.  


We adopt a 12-layer chain-topology LeNet  model~\cite{lecun1998gradient, nishio2019client}, which consists of six convolution (CONV) layers,  three max-pooling (POOL) layers, and three fully-connected (FC) layers. The detailed model structure and model parameters are shown in Table~\ref{DNN model parameters}. The whole model has around 4.3 million parameters, and each parameter is quantized into 32\;bits, leading to a data size of about 16.49\;MB. The smashed data and its gradient are also quantized into 32\;bits.  There are 5 devices in a cluster. The mini-batch size is set to 16.


\begin{figure}[t]
	\centering
	\renewcommand{\figurename}{Fig.}	
	\begin{subfigure}[Number of training rounds]{
			\label{fig:training_round_non_IID_with_FL}
			\includegraphics[width=0.38\textwidth]{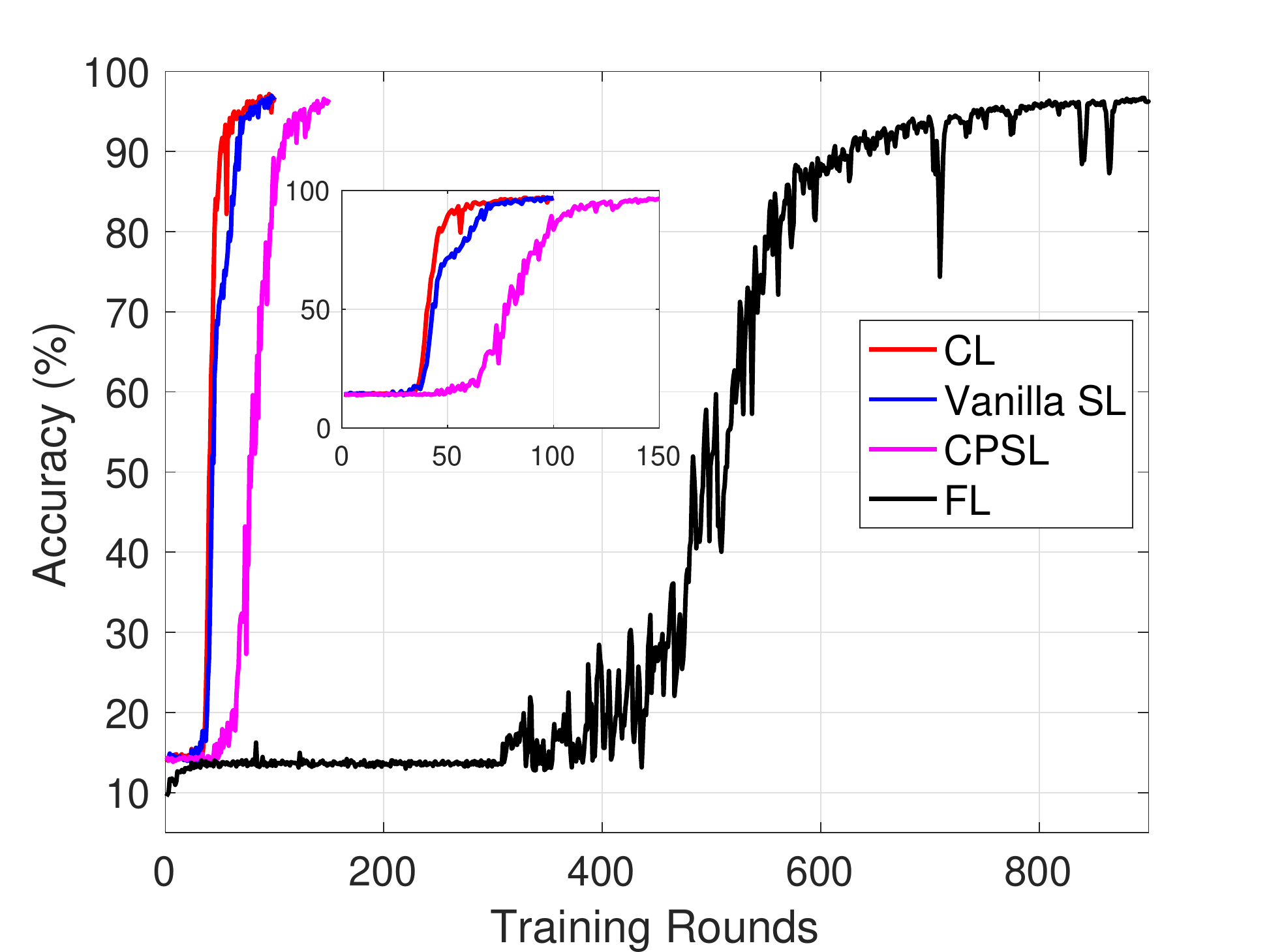}}
	\end{subfigure}
	~
	\begin{subfigure}[Training latency]{
			\label{fig:convergence_time}
			\includegraphics[width=0.38\textwidth]{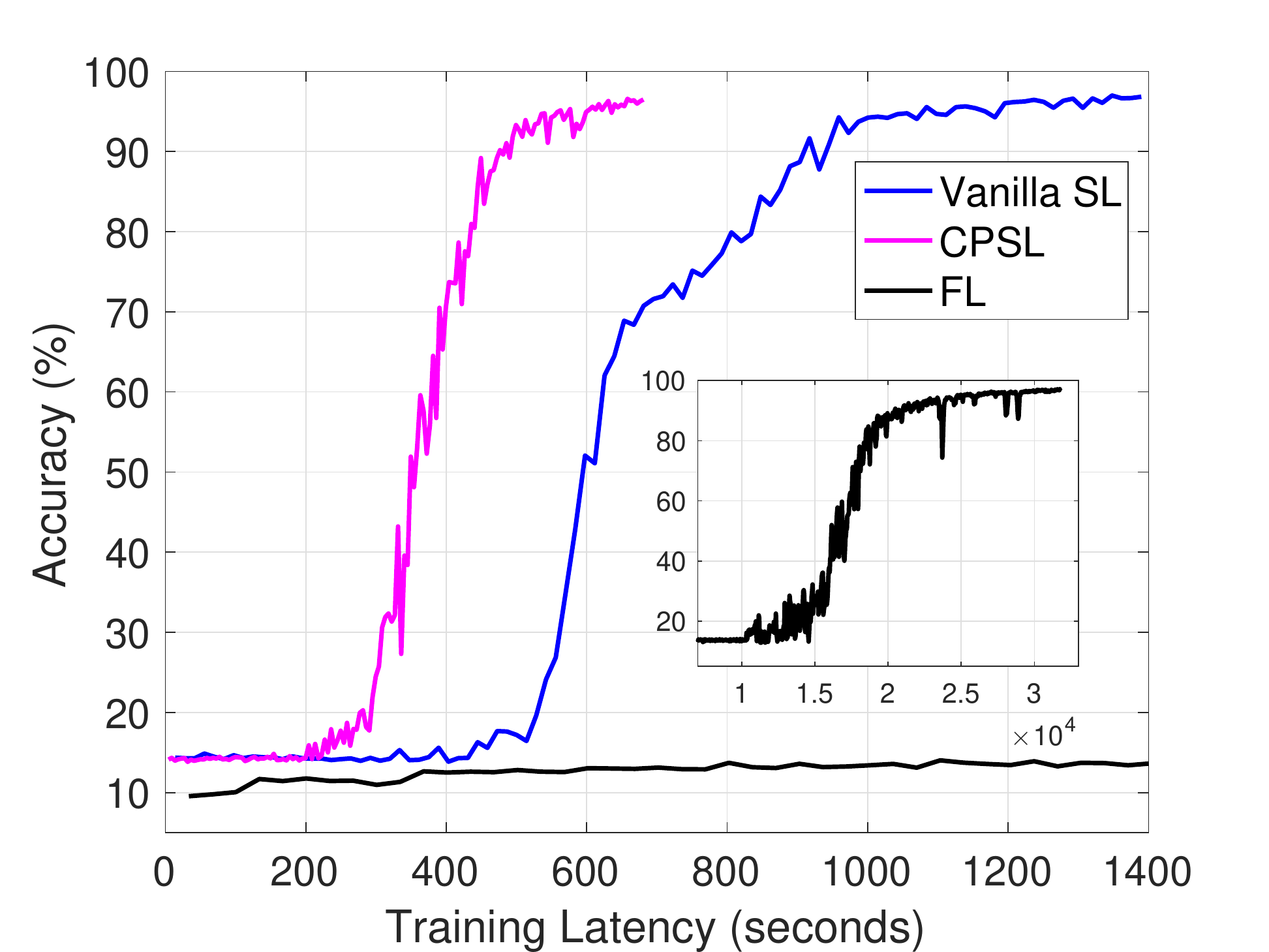}}
	\end{subfigure}
	\caption{Training performance comparison among different schemes over non-IID data distribution. }
	\label{Fig:convergence_performance_comparsion}
		\vspace{-0.7cm}
\end{figure}

\subsection{Performance Evaluation of the Proposed CPSL Scheme}\label{subsec:PSL Algorithm}
To better elaborate the performance evaluation of  the proposed CPSL algorithm, we consider that devices are \emph{identical} in terms of computing and communication capabilities. The computing capabilities and the received signal to noise (SNR) threshold of each device are set to 0.5$\times 10^9$\;cycles/s and 17\;dB, respectively. Given the selected cut layer, the data size of the device-side model is 0.67\;MB. The data sizes of smashed data and its gradient for one data sample are 18\;KB and 9\;KB, respectively. The FP computation workloads of the device-side model and the server-side model are 5.6\;MFlops and 86.01\;MFlops, respectively. The computation workloads of FP and BP processes are assumed to be the same.

We compare the proposed CPSL algorithm with the following benchmark schemes: (1) \emph{centralized learning (CL)}, which can achieve the optimal model convergence and accuracy. In practice, it is difficult to implement due to privacy leakage concerns and prohibitive communication overhead; (2) \emph{vanilla SL}, where the AI model is sequentially trained across all the devices~\cite{gupta2018distributed}; and (3) \emph{FL}, where all the devices train the shared model locally, and then send the trained models to the edge server for new model aggregation in each iteration~\cite{bonawitz2019towards}. For fair performance comparison, all the schemes adopt the same model initialization. The learning rates of CL, vanilla SL, and FL schemes are optimized, which are set to 0.05, 0.05, and 0.1, respectively.

\subsubsection{Training Performance}
Figure~\ref{fig:training_round_non_IID_with_FL} shows the training performance of the proposed scheme and all the benchmarks.  Several important observations can be obtained.  Firstly, the proposed scheme can achieve nearly the same accuracy as CL and SL, which validates its remarkable performance, at the cost of more training rounds. This is because device-side model aggregation in each cluster slows down the model convergence. Specifically, the proposed scheme takes about twice training rounds to converge. Secondly, FL converges much slower than other algorithms due to model aggregation among a large number of devices. In addition, due to a heavy device computation workload, FL takes extremely long training latency before  convergence.

Since the per-round training latency of different schemes is different, we further evaluate the overall training latency in Fig.~\ref{fig:convergence_time}. The overall training latency is the product of the per-round training latency and the number of training rounds. The proposed scheme takes a shorter training latency than the SL to reach convergence. Specifically, the time consumed by the proposed scheme is about 600 seconds, while that by  SL is about 1,400 seconds.  The reason is that the per-round training latency of the proposed scheme is much smaller than that of the SL.
The per-round training latency of the CPSL, SL, and FL are 3.78\;seconds, 13.90\;seconds, and 33.43\;seconds, respectively.

\begin{figure}[t]
	\centering
	\renewcommand{\figurename}{Fig.}	
	\begin{subfigure}[MNIST dataset]{
			\label{fig:different_clients}
			\includegraphics[width=0.38\textwidth]{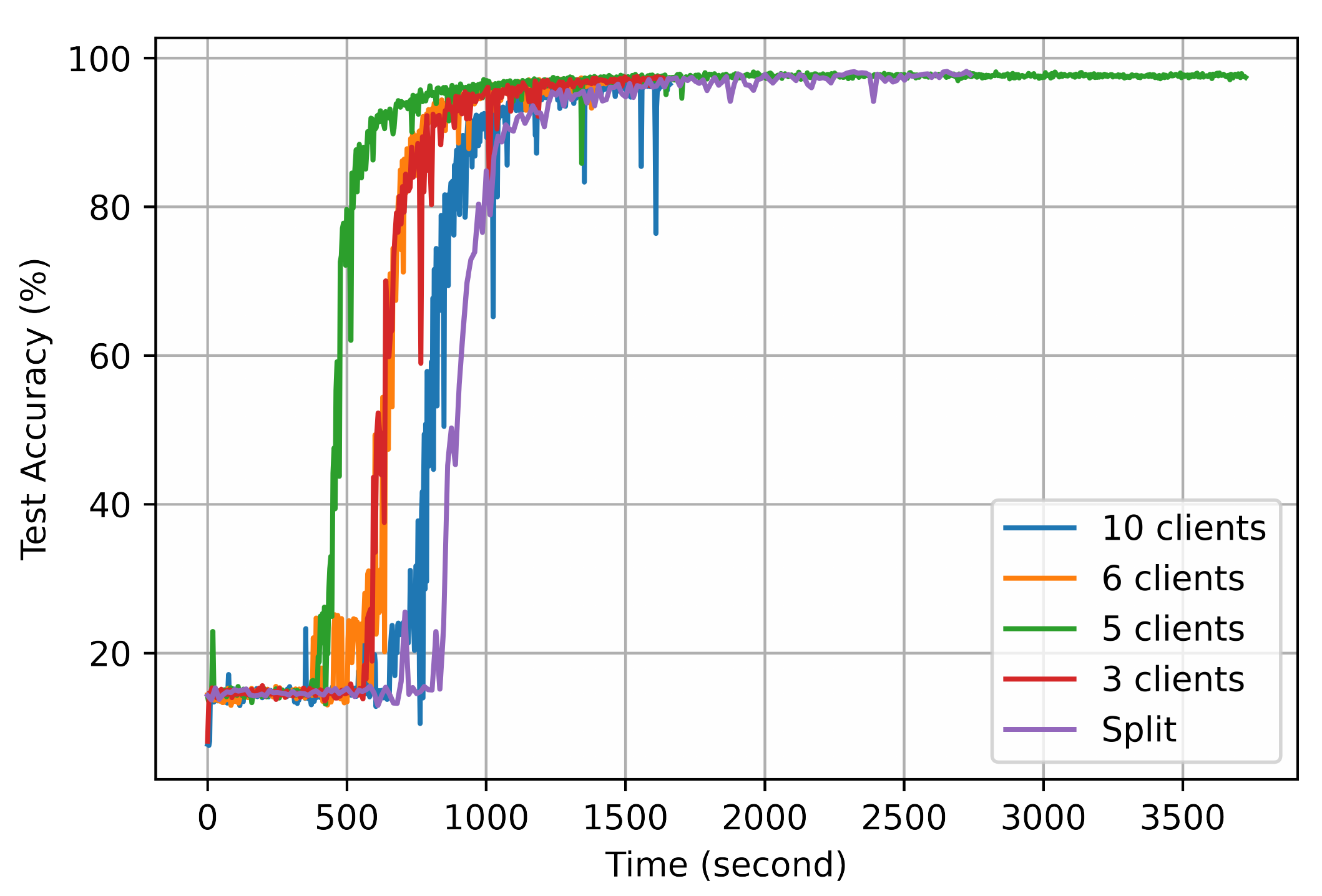}}
	\end{subfigure}
	~
	\begin{subfigure}[Fashion-MNIST dataset]{
			\label{fig:FMNIST-3}
			\includegraphics[width=0.38\textwidth]{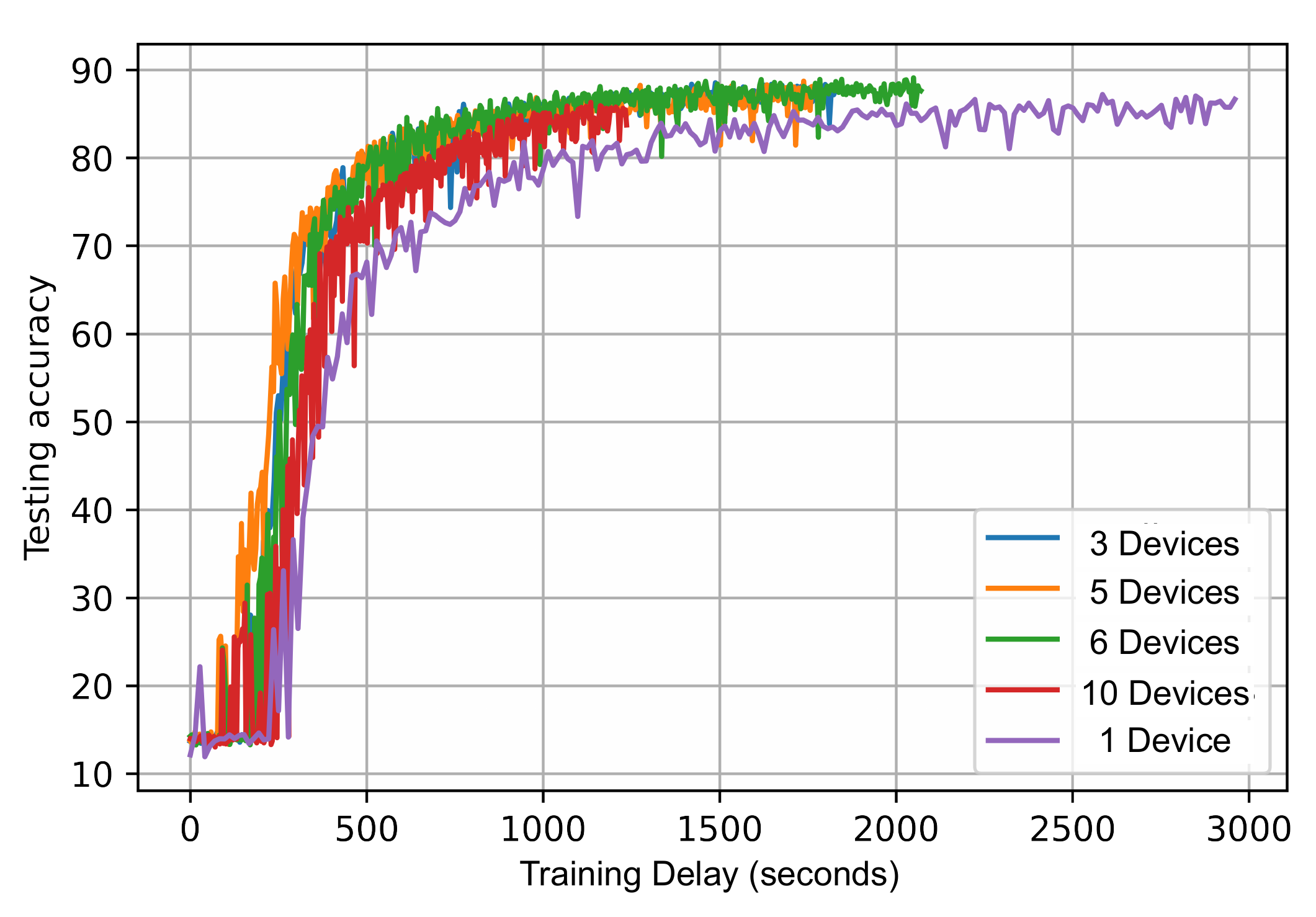}}
	\end{subfigure}
	\caption{Overall training latency  with respect to the number of devices in a cluster. }
	\label{Fig:performance_comparsion_dataset}
	\vspace{-0.7cm}
\end{figure}








\subsubsection{Impact of Cluster Size}
Figure~\ref{fig:different_clients} compares the performance with respect to different numbers of devices $N_m$ in a cluster. Several key observations can be obtained. Firstly, the number of devices in a cluster affects the training latency to achieve convergence. Specifically, the proposed scheme with 5 devices in a cluster has the lowest training latency. Secondly, the proposed scheme for different numbers of devices from 3 to 10 can converge faster than  SL because device-side models are trained in parallel in the proposed scheme. Thirdly, all the schemes achieve nearly the same accuracy at the end of the training process. This indicates that the proposed scheme does not incur any accuracy loss while reducing the training latency. A similar simulation is conducted on the Fashion-MNIST dataset, with results shown in Fig.~\ref{fig:FMNIST-3}. It can be seen that the proposed scheme effectively reduces overall training latency as compared with SL while preserving model accuracy.

%


\subsection{Performance Evaluation of the Proposed Resource Management Algorithm}\label{subsec: DOPA}
We evaluate the performance of the proposed resource management algorithm by taking device heterogeneity and network dynamics into account. The mean values of device computing capability and the SNR of the received signal are randomly drawn from uniform distributions within $[0.1, 1]\times 10^9$\;cycles/s and  $[5, 30]$\;dB, respectively.  Standard deviation $\sigma_f$ and $\sigma_h$ are set to $0.05\times 10^9$\;cycles/s and 2\;dB, respectively.
\begin{figure}[t]
	\renewcommand{\figurename}{Fig.}
	\centering
	\includegraphics[width=0.38\textwidth]{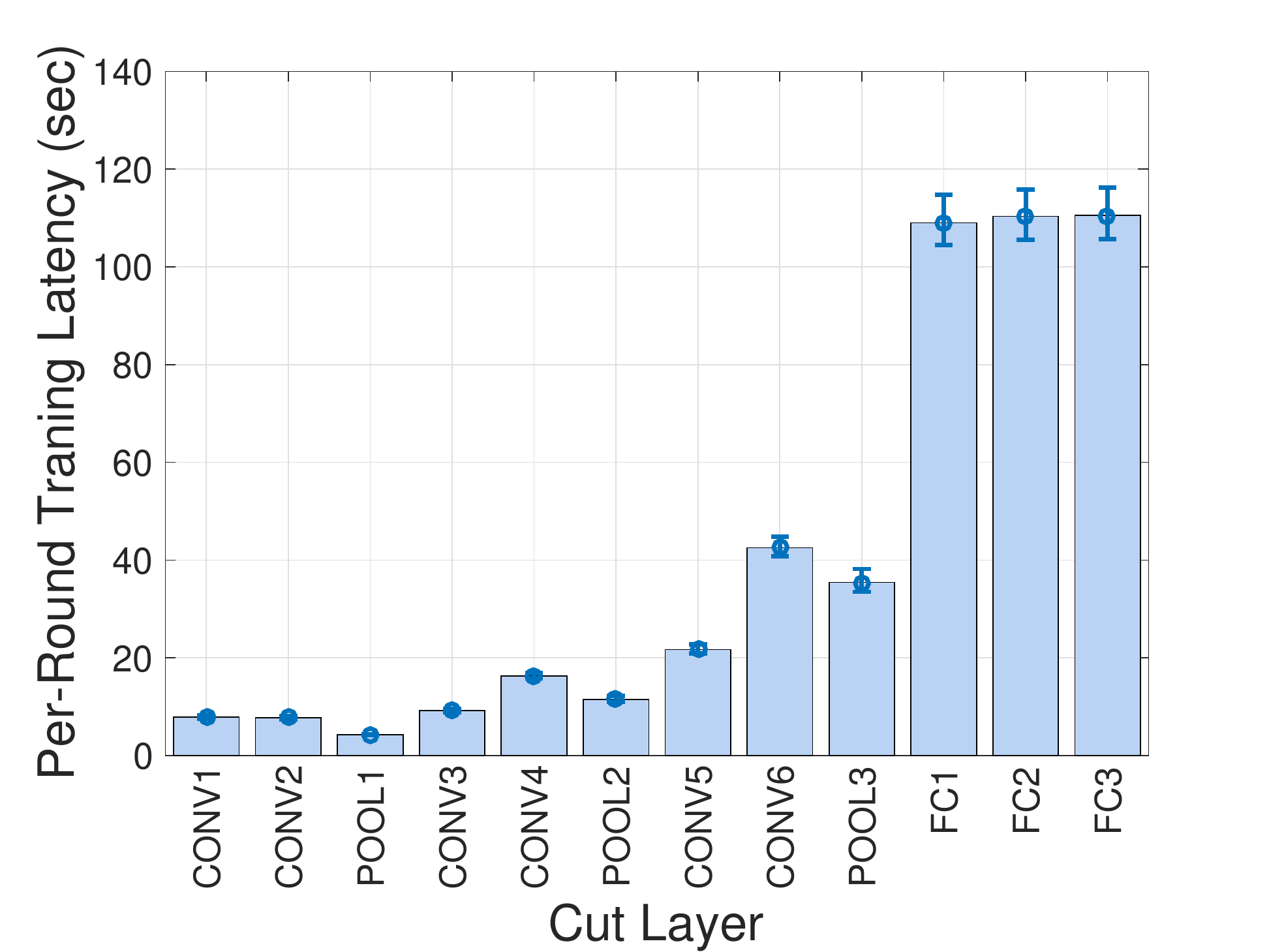}
	\caption{Per-round training latency with respect to different cut layers. Error bars show the 95 percentile performance.}
	\label{fig:latency_proposed}
	\vspace{-0.5cm}
\end{figure}

Figure~\ref{fig:latency_proposed} presents per-round training latency with respect to different cut layers  over 300 simulation runs. The POOL1 layer achieves the minimum average per-round training latency, which is selected as the optimal cut layer. This is because this layer results in a small amount of communication overhead and balances the computation workload between the device and the edge server.

\begin{figure}[t]
	\centering
	\renewcommand{\figurename}{Fig.}	
	\begin{subfigure}[Convergence]{
			\label{fig:convergence}
			\includegraphics[width=0.38\textwidth]{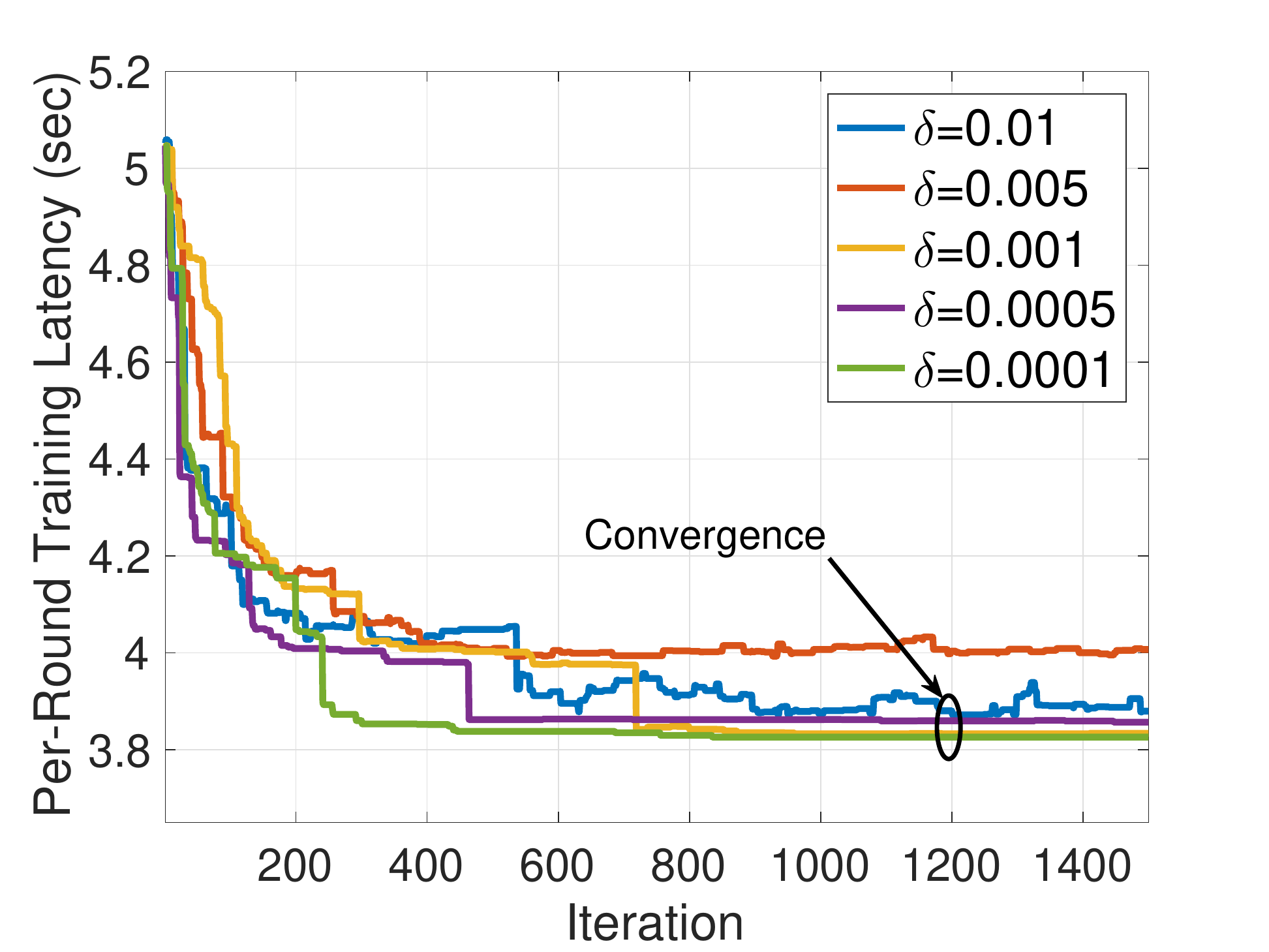}}
	\end{subfigure}
	~
	\begin{subfigure}[Training latency]{
			\label{fig:training_latency_proposed}
			\includegraphics[width=0.38\textwidth]{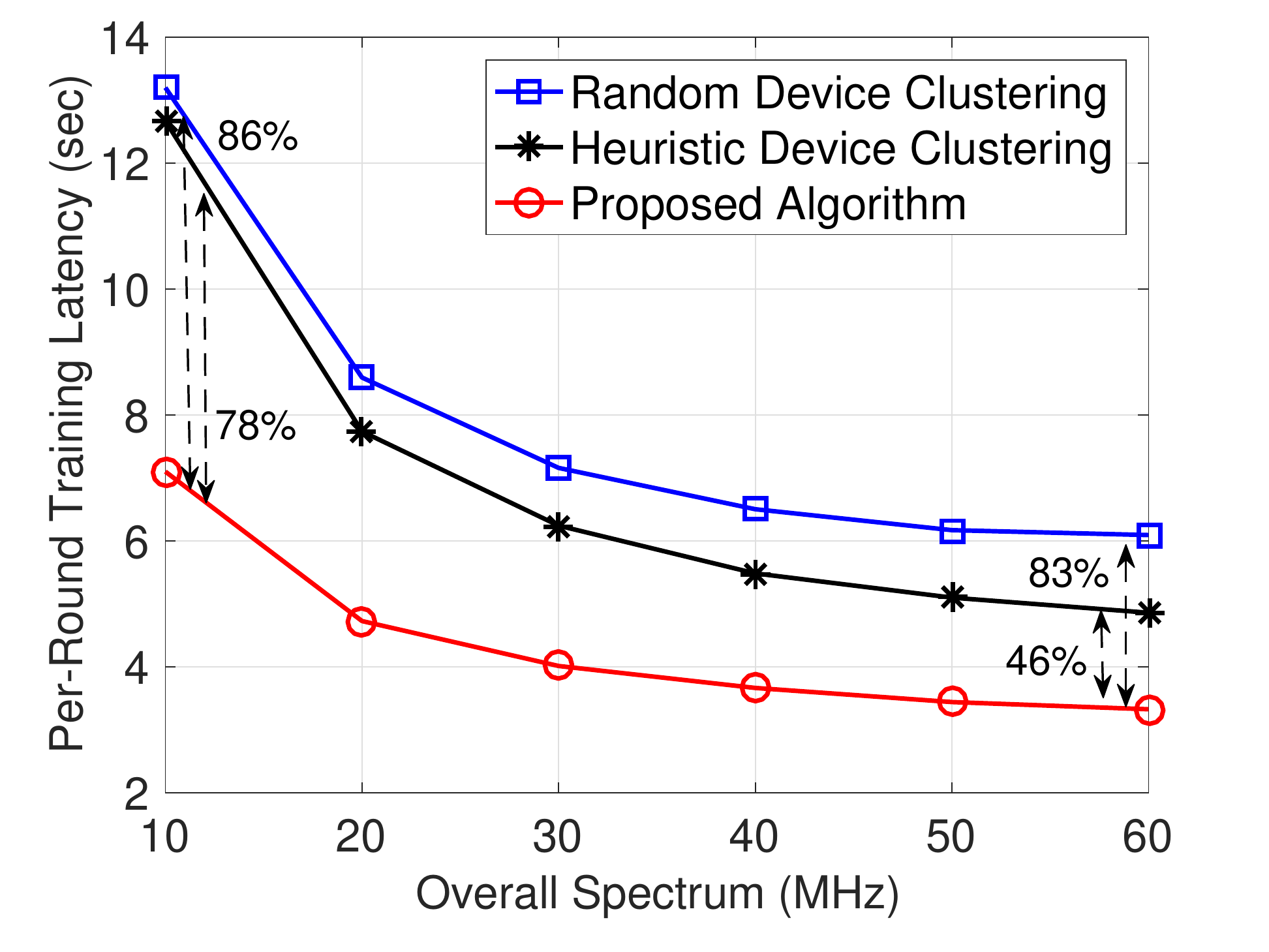}}
	\end{subfigure}
	\caption{Performance comparison among the proposed algorithm and benchmarks. }
	\vspace{-0.9cm}
\end{figure}

 Figure~\ref{fig:convergence} shows the convergence process of the proposed resource management algorithm. When smooth factor $\delta =0.0001$, the proposed algorithm converges after about 1,000 iterations, although it stays in several local optima for a while before identifying the global optimum. However, further increasing the value of smooth factor (e.g., $\delta =0.01$) may impede the identification of global optimum and result in the convergence to inferior solutions.

Figure~\ref{fig:training_latency_proposed} compares the proposed algorithm with two benchmarks: (1) \emph{heuristic device clustering algorithm}, where devices with similar computing capabilities are partitioned into clusters; and (2) \emph{random device clustering algorithm}, which partitions devices into random clusters. We see that the proposed algorithm can significantly reduce per-round training latency as compared with the benchmarks, because device clustering and radio spectrum allocation are optimized. Specifically, the proposed algorithm reduces the training latency on average by 80.1\% and 56.9\% as compared with the heuristic and random benchmarks, respectively. In addition, the performance gain achieved in spectrum-limited scenarios (e.g., 10\;MHz) is  higher than that in the scenarios with more radio spectrum resources (e.g., 60\;MHz), highlighting the importance of the proposed resource management algorithm in alleviating the straggler effect of CPSL in spectrum-limited wireless networks.

\section{Conclusion}\label{sec:conclusions}
In this paper, we have investigated a training latency reduction problem in SL over wireless networks. We have proposed the CPSL scheme which introduces parallelism to reduce training latency. Furthermore, we have proposed a two-timescale resource management algorithm for the CPSL to minimize the training latency in wireless networks by taking network dynamics and device heterogeneity into account.  Extensive simulation results validate the effectiveness of the proposed solutions in reducing training latency as compared with the existing SL and FL schemes. Due to  low communication overhead, device computation workload, and training latency, the CPSL scheme can be applied to facilitate AI model training in spectrum-limited wireless networks with a large number of resource-constrained IoT devices. For future work, we will investigate the impact of device mobility on SL performance.



\bibliographystyle{IEEEtran}
\bibliography{Trans8_reference}

\end{document}